# First-principles studies of multiferroic and magnetoelectric materials

**Yue-Wen Fang [1], Hang-Chen Ding [1], Wen-Yi Tong[1], Wan-Jiao Zhu[1], Xin Shen[1], Shi-Jing Gong[1*], Xian-Gang Wan [2], Chun-Gang Duan [1,3,*]**

[1] Key Laboratory of Polar Materials and Devices, Ministry of Education, East China Normal University, Shanghai 200062, China;

[2] Department of Physics and National Laboratory of Solid State Microstructures, Nanjing University, Nanjing 210093, China;

[3] National Laboratory for Infrared Physics, Chinese Academy of Sciences, Shanghai 200083, China

[*]Corresponding authors (email: sjgong@ee.ecnu.edu.cn, wxbdcg@gmail.com)

**Abstract:** Multiferroics are materials where two or more ferroic orders coexist owing to the interplay between spin, charge, lattice and orbital degrees of freedom. The explosive expansion of multiferroics literature in recent years demonstrates the fast growing interest in this field. In these studies, the first-principles calculation has played a pioneer role in the experiment explanation, mechanism discovery and prediction of novel multiferroics or magnetoelectric materials. In this review, we discuss, by no means comprehensively, the extensive applications and successful achievements of first-principles approach in the study of multiferroicity, magnetoelectric effect and tunnel junctions. In particular, we introduce some our recently developed methods, e.g., the orbital selective external potential (OSEP) method, which prove to be powerful tools in the finding of mechanisms responsible for the intriguing phenomena occurred in multiferroics or magnetoelectric materials. We also summarize first-principles studies on three types of electric control of magnetism, which is the common goal of both spintronics and multiferroics. Our review offers in depth understanding on the origin of ferroelectricity in transition metal oxides, and the coexistence of ferroelectricity and ordered magnetism, and might be helpful to explore novel multiferroic or magnetoelectric materials in the future.

## 1 Introduction

In the 1890s, famous French scientist Pierre Curie, from symmetry point of view, asserted that non-moving crystals could demonstrate magnetoelectric effect [1]. We have no clue that how he could make such penetrative conclusion at his era, considering that the origins of magnetism and ferroelectricity could only be clarified after the birth of quantum mechanics, and are still in debates in many cases by now. Nevertheless, apparently he had opened a new world for the material science. 100 years later, in the celebration of centennial of Curie's idea, the term multiferroic, i.e., two or more of the primary ferroic properties are united in the same phase, was coined by Schmid [2]. After another 20 years, the research on magnetoelectric and multiferroic materials is already a big and important branch of the condensed matter physics. For the concepts and phylogeny of magnetoelectric and multiferroic materials, one can refer to a series of reviews [3-12].

In the last twenty years, it is obvious that ab initio, or first-principles calculation has played a very important role in the revival of magnetoelectric and multiferroic research. It is therefore the aim of the current review to give an introduction of the application of first-principles calculation in the study of magnetoelectric effect and multiferroic materials. Indeed, thousands of papers related to the first-principles calculation on the magnetoelectric or multiferroic properties have occurred in the past decades. It is definitely impossible to give a comprehensive review to introduce them all. Fortunately, there are already some nice reviews in this subject [13-15]. We then focus our review on the recent important progress in the first-principles studies of magnetoelectric and multiferroic materials.

This review is organized as follows. We first give a brief overview of first-principles electronic structure calculations. We emphasize the aspects of first-principles calculations relevant to the study of magnetoelectric and multiferroic materials. Particularly, we will introduce some newly developed computational methods, which are proved to be helpful to clarify the mechanisms of the ferroelectric, magnetic and magnetoelectric effect in the system. Then we give some recent representative examples of the successful applications of first-principles calculations in the explanation and prediction of magnetoelectric or multiferroic phenomena. Both single-phase and composite multiferroics are involved. As the concept of electric-field control of magnetism is indeed the core of the magnetoelectric and multiferroic research, we will also show some progresses towards this direction, e.g., the ab initio study of Rashba effect. Finally we give a summary and perspective of first-principles calculations in the field of magnetoelectric and multiferroic materials.

## 2 First-principles calculations: advantages and limitations

Nowadays, relying on evolution of numerous numerical algorithms and modern computers, computational physics gradually bridges the gap of theoretical physics and experimental physics. Among the computational approaches, first-principles calculations based on solving of true fundamental Hamiltonian of the system build up the pillars of quantum many-body research especially in the field of multiferroic materials. As solving Schrödinger equation for many-body system is a challenging work even for high performance supercomputing center, it's unavoidable to introduce some approximations to curtail the results of ab initio calculations. However, these approximations will not, in general, introduce adjustable physical parameters into system except for making computational tasks easier.

The application of first-principles calculation has been more and more extensive, and will continuously grow exponentially. In the research field of materials, physics, chemistry, and even biology, first-principles study has gradually become an important research technique, which covers both macroscopic and microscopic scales. As it is too difficult and lengthy to make a detailed and comprehensive introduction about the application of first-principles calculation, we will only make a brief introduction here, focusing on the part of magnetoelectric and multiferroic studies.

### 2.1 Why first-principles calculations?

As mentioned above, first-principles calculations are in principle parameter free calculations. Compared with traditional analytic method, many complicated and time-consuming calculations can be carried out from first-principles, which would largely improve the accuracy and reliability of results. On the other side, contrary to the modern experimental method, which is based on precise measurement using highly sensitive and expensive instruments and devices, and largely depends on the quality of samples, the first-principles study is almost cost-free and can be easily applied in many materials and systems. This will inevitably shorten the period of research to a great extent.

Particularly, the first-principles calculation is powerful in the prediction of physical properties of materials under various extreme conditions. For examples, the density functional theory (DFT) in the standard formulation is a zero-temperature method [16], and thus it is naturally used to deal with the problem under very low temperature, which will be difficult and expensive to realize in experiments. In addition, systems experiencing extreme high-temperature can be simulated by first-principles molecular dynamics (FPMD) [17]. Besides the conditions under extreme temperature, the first-principles calculations allow detailed studies of systems under high (hydrostatic) pressures [18], large epitaxial strain [19], and strong (electric, magnetic and optical) fields [20, 21], or with heavy doping [22], etc. It is also relatively easy to study low-dimensional systems, for instance quantum well [23], quantum wire [24], quantum dot [25], perfect surface [26] and interface [27].

As a more practical method than analytic one, as well as a much cheaper tool compared with experiment, first-principles calculation has been successfully applied to predict hypothetical or unknown structures and their properties, some of which have already been proved experimentally [28]. Indeed, in the last three decades, first-principles calculations have gradually evolved from mere aids in supporting experimental data and interpreting physical phenomena observed from diverse experimental techniques or obtained from theoretical research, as well as revealing the mechanism of them [29], to powerful tools for the design of novel materials.

It is foreseeable that with the theoretical developments and increases in the calculation capability, first-principles calculations will reach unprecedented levels of accuracy in treating complicated systems at the microscopic level [29] and play a more and more important role in the research of physics, chemistry, materials science, etc.

2.2 What can first-principles calculations do?

For the system given atomic positions and element types, the total energy can be easily obtained by solving Kohn-Sham equation using DFT based self-consistent calculation. Due to the fact that the stable structures in nature possess the lowest total energy, structure optimization can be realized during the process of calculating the minimum energy. In the past few decades, the accurate first-principles total-energy calculation has been widely used for investigation of phase transition and phase stability. Classical cases include studies of the relative energetics of three different silica polymorphs [30], the stability of $Li_xCoO_2$ for $x$ ranging from 0 to 1 [31], and the transition from the *Cm* post-perovskite phase to *Pbnm*-perovskite phase in $MgSiO_3$ [32]. It has also been used to explore the phase transition in multiferroics, such as strain-induced isosymmetric phase transition in $BiFeO_3$ (BFO) [33], structure instability of four cubic perovskite $BiMO_3$ compounds (M = Al, Ga, In, and Sc) [34], and C-type to G-type antiferromagnetic phase of tetragonal-like (T-like) BFO [35].

For dynamic simulations and relaxations of first-principles calculation, density of states (DOS), as well as the band structure are available, which are widely used to explore the electronic properties in compounds, e.g., perovskite compounds [36-38], wurtzite-type compounds [39-41]. In general, the calculation of band structure demonstrates the nature of the band gap, the position of the valence (conduction) band edge, and the features at high symmetry points, based on which, the metal-insulator transitions [42, 43], and direct-indirect band gap transition [44] can be researched. For the first-principles DOS calculation, except for the conventional total DOS, integrated DOS, and site projected DOS, the DOS between two spin states, as well as the energy resolved magnetization density in the $x$, $y$ and $z$ direction is also available for spin-polarized case. The overlapping DOS indicate bonding between different states and inflect the degree of hybridization [45, 46].

The elastic modulus, strength are important mechanical properties of materials. The elastic modulus contains many types. The three primary ones are the tensile modulus (or the Young's modulus), the shear modulus, and the bulk modulus. Due to the good agreement with experimental data, the first-principles calculations of the elastic modulus have already been an effective and significant method to explore the elastic properties of materials [47-49]. The theoretical strength of materials describes the upper limit to withstand an applied stress without failure, including the compressive strength, the tensile strength and the shear strength. The first-principles strength calculation has already been the hotspot for the research field of computational materials science and has been widely used for metals [50, 51], diamond-type materials [52, 53], and carbon-nanostructures [54, 55]. With the application of first-principles calculation, the electronic structures and mechanical properties can be calculated simultaneously, which will contribute to the explanation for the relationship between structures and mechanical properties, as well as the theoretical guidance in the development of new materials.

The physical properties of materials mainly include three aspects, i.e., magnetic, electric and optical properties. As these properties of materials are directly related to the crystal and electronic structures, first-principles calculations of physical properties have been proved to be reliable by comparison with experiments, and consequently have been widely used.

Magnetism is a phenomenon by which materials assert an attractive or repulsive force on others. The nature and magnitude of the atomic magnetic moments determine the magnetism of materials in large part. Using first-principles calculation, the spin and orbital contributions to the magnetic moments can be available [56, 57], based on which, the origin of the magnetism [58-60] and the way to enhance the magnetism [61] can also be investigated. For the research of magnetism using first-principles calculation, the magnetocrystalline anisotropy energy (MAE) is one of the important fields. The representative work is Daalderop et al's [62] research of iron, cobalt, and nickel, Wang et al.'s [63, 64] state tracking approximation and surface magnetocrystalline anisotropy analysis, tetragonal FeCo alloys with giant MAE predicted by Burkert et al. [65], and the effects of an external electric field on the MAE of transition-metal films [66, 67]. In addition, the analysis of magnetic ordering and spin exchange interaction has been carried out using first-principles calculation in multiferroics [68, 69], low-dimension materials [70, 71] and other compounds [72].

Electricity is a set of physical phenomena associated with the presence and flow of electric charge. For insulators, what we concerned is the dielectricity, which describes the ability of the system to be polarized under an external electric field. For conductors or ultrathin insulating films where electron tunneling could occur, we are able to study their transport properties using first-principles calculations [73]. Whereas for dielectric materials, the main application of first-principles calculation in dielectricity is the research of polarization. The enhancement of polarization in heteroepitaxially constrained thin films of the multiferroic BFO, reported by Wang et al. [74], and Fu et al's [75] work about a large piezoelectric response driven by polarization rotation induced by an external electric field are two of the representatives among them. First-principles study of spontaneous polarization is extensively applied in multiferroics, such as BFO [38, 76], $Bi_2FeCrO_6$ [77], $SrMnO_3$ [78], and ferroelectrics, such as $BaTiO_3$ (BTO) [79], $PbTiO_3$ (PTO) [80]. Especially, the study of polarization rotation is widely used to design piezoelectric materials, and explain the origin of electromechanical response [81, 82]. Furthermore, the study of ferroelectric domain wall [83, 84] is also available using first-principles calculations.

Optics is the branch of physics which involves the properties of light and its interactions with matter. According to the response of dielectric polarization to the electric field of the light, optics has been divided into two parts: linear optics and nonlinear optics. Linear and nonlinear optical techniques are particularly important as tools for providing complementary information on crystallographic, electronic and magnetic properties and studying the coexistence and interactions of magnetic and electric order. As the fact that neither the standard DFT [85] nor the GW approximation (GWA) [86] allows for a correct evaluation of optical spectra or other charge-neutral excitations, using first-principles calculation in optical response is much later than its application in magnetic and electronic properties [87]. With the development of single- and many-particle Green's function so far, first-principles study has been extensively carried out to explore linear and nonlinear optics, such as real and imaginary part of dielectric functions [45, 88], absorption, reflectivity, refractive spectrum [89-93], magneto-optical Faraday and Kerr spectra [94-96], second-harmonic generation susceptibilities [97-100], and obtained excellent agreement with experimental measurements. Recently, Tong et al. [101] extended the first-principles optical calculation to the spin-dependent case, which will be helpful to study the optical properties of magnetic systems.

First-principles calculations are especially useful in studying low-dimensional systems, which are the frontiers of condensed matter physics. Surface—be it of a metal, an ionic or covalent solid, or a semiconductor—is a form of matter with its own physics. It is similar to bulk materials over many aspects while its particular structure of low dimension brings numerous differences which make surface science more glamorous. Actually, a surface can be regarded as an interface between a material and vacuum. The interface physics tends to be more complicated and fruitful when involving several different materials because it is likely to bring fascinating phenomena, e.g., concomitant interface two dimensional electron gas (2DEG), interface ferromagnetism [102], phase transformation [103], interface polarization pinning [104] or interface effect modified MAE [105], etc. Consequently, interfaces are usually expected to be more functional as the famous phrase goes coined by Nobel laureate Herbert Kroemer, the interface is the device [106]. However, some quantitative investigations for interfaces and surfaces are unreliable or difficult to measure in experiments technically. First-principles calculations can help to modify or confirm these kind of results, which make up the deficiency of experiments. Up till now, the first-principles studies have managed to calculate the current-voltage (*I-V*) characteristics of a molecular device concerning interfaces [107], to predict the critical thickness for the ferroelectricity affected by a realistic ferroelectric-electrode interface [108, 109] and magnetoelectric effect in multilayers [110], as well as to investigate the interface electrochemistry in conversion reactions [111], etc. Additionally, when it comes to nanotechnology and spintronics in the last two decades concerning many low dimensional materials, e.g., 2D black phosphorus [112], carbon nanotubes [113], graphene [114], silicene [115], topological insulator nanostructures [116] etc., the overwhelming contribution from first-principles calculations should never be ignored because of thousands of prospective studies involving fundamental physics of transport properties [117, 118] and thermal stability [119, 120], material characterization [121, 122] and devices design [123, 124] have been (or being) proved (or fabricated), and further realized into novel nanodevices in the future.

As is well known, classical molecular dynamics simulations describe the process of resolving Newton's equations of motion for many-body systems over a period of time. As one of the most powerful modern computational methods, classical molecular dynamics study has been applied into ultra-precision manufacturing [125], (soft) condensed matter physics [126-128], genetic engineering [129], medicine [130], etc. However, its intrinsic dependence of predefined potentials and force fields established by empirical physical data or independent electronic structure calculations not only bring suitable approximation for atomic interactions, but also is a conspicuous shortage owing to the terrible portability of potentials. For an unexplored matters or materials, classical molecular dynamics is frustrating for the absence of a set of certified self-consistent potentials. To overcome this problem, a technique reigning classical molecular dynamics and general first-principles study was created as FPMD, as well as the conventional name Car-Parrinello molecular dynamics (CPMD) naming after two Italian physicists Car and Parrinello [131] who proposed this set of underlying theory in 1985.

Over the last thirty years, FPMD has made a substantial progress and attracted many eyes in computational physics, computational chemistry and computational biology. As said by physicist Hutter, one of the developers of CPMD and CP2K program packages, its applications are instrumental in fields as diverse as enzymatic catalysis and the study of the interior of planets [132]. Though FPMD is versatile, the underlying idea of every FPMD is to compute the forces exerting on nuclei relying on interaction potentials from the on-the-fly electronic structure calculations as the molecular dynamics trajectory is formed. FPMD technique within the framework of DFT open ways for us to observe the time-dependent and temperature-dependent physical behaviors dynamically. Nowadays, the track of FPMD can be found in phase transitions [133], surface and interface chemistry [134], liquids properties investigations [135] and other physical or chemical research [136, 137]. In a word, FPMD is not a mechanically simple combination of classical dynamics and general first-principles theory, it has been cooked into a state-of-the-art theoretical tool actually.

## 2.3 Limitations of first-principles calculations

First-principles calculation, of course, is not a remedy for every problem in the research of condensed matter physics. Like any other scientific tools, it has its own limitations. For the magnetoelectric and multiferroic materials as we concerned in this review, the limitations can be summarized as follows.

The first and also the most serious one is the accuracy of total energy calculations. As we know, in the application of DFT to the first-principles calculation, we have to choose approximate density functionals to treat electronic exchange-correlation energy. Among the approximate density functionals, local density approximation (LDA) and local spin-density approximation (LSDA), whose philosophy can be dated back to Thomas–Fermi–Dirac theory, is probably the most widely employed within first-principles calculation. However, LDA usually overestimates the binding energy of molecules and the cohesive energy of solids due to lack of cancelation of the self-interaction, consequently tends to underestimate bond lengths especially for systems involving hydrogen bonds or van der Waals closed shell interactions. Besides, accurate band gap is difficult to obtain in LDA for semiconductors and strongly correlated systems such as transition metal oxides CoO [138]. On account of these limitations, many other exchange-correlation functionals were proposed. Generalized gradient approximation (GGA) is the most preferred one in practice since it improves binding energies, bond lengths and computational efficiency, but some problems still exist such as unreasonable gap energy. For the multiferroic materials we concerned, which inevitably involve magnetic transition-metal ions, the ordinary first-principles calculations sometimes may provide unreliable results. Of course, more accurate many-body methods, like GW, dynamical mean field theory (DMFT), or more comprehensive density functional approximations, like hybrid GGA, meta-GGA, can be adopted to overcome these problems to some extent [139-142]. These methods, unfortunately, at present are far from being practical for the complex structures of interest.

Second, as the devices scale down into nanoscale, first-principles study enables us to design and test novel nanodevices, molecular devices and spintronics devices. Whereas, first-principles study is a double edged sword. Once applied to investigate large scale systems containing more than thousands of atoms or more, it is no longer an efficient way due to the rapid increase of computing time. This sets a severe limitation on the phenomenon that first-principles calculation could handle. For instance, the ferroelectric domain and magnetic domain generally involve groups of atoms, hence hard to be studied by first-principles calculation.

Lastly but not least, first-principles approaches to uncover various properties of the systems are still limited. Basically we only have confidence on predicting some ground state properties of the system. As to the properties of non-periodic, or non-equilibrium systems, there is still a long way to go. Therefore, work toward a method that takes advantage of current knowledge, experience with existing methods as GW method for excited state systems [143] and kinetic energy density functionals for non-periodic systems [144] may lead to real progress.

## 2.4 Recent progress in developing first-principles tools

As stated above, to fill the gap between the basic DFT calculations and the prediction of demanded properties of the system, one must develop appropriate theory. For a periodic system, the polarization is a lattice rather than a vector, and the polarization lattice can be calculated by quantum mechanical electronic structure methods [145]. In fact, with the development of Berry phase theory, the ferroelectric polarization can also be obtained in DFT calculation [146]. Berry phase theory rigorously defines the spontaneous polarization of a periodic solid and provides a route for its computation in electronic structure codes. Another recent progress in DFT for periodic solids especially for insulators is attributed to Thonhauser et al [147]. They derived an expression for the orbital magnetization of a periodic insulator by working in the Wannier representation, which is a significant contribution to a more general theory of orbital magnetization. The development of relevant theories and numerically calculation of specific physical quantities is always an important theme in the history of first-principles calculations.

Another direction of the development of the first-principles calculations is to make compromise between the complicated many-body calculation and the simple yet less accurate single-particle calculations when handling the strong correlated systems. Among these efforts, the LDA+*U* is perhaps the most successful one [148, 149]. In fact, orbital-sensitive potentials are proved to be effective in solving many subtle problems in transition-metal oxides. Recently, Wan and co-workers [150, 151] developed a new approach, orbital selective external potential (OSEP) method, which was previously called constrain orbital-hybridization method to effectively investigate the origins or mechanisms of the intriguing properties of the systems we are interested in. The spirit of the OSEP approach is to introduce a special external potential. Different from the realistic external potential, this potential is orbital sensitive, i.e., only certain appointed orbital can feel it. Though this orbital sensitive potential is originally proposed for theoretical purposes, indeed such potential could exist in nature. Specifically, a projector operator $|inlm\sigma\rangle\langle inlm\sigma|$ is defined, which only allow the external potential $V_{ext}$ influence the specific atomic orbital $|inlm\sigma\rangle$. Here *i* denotes the atomic site, and *n*, *l*, *m*, $\sigma$ are the main quantum number, orbital quantum number, magnetic quantum number and spin index, respectively. The new Hamiltonian can be written as:

$$H^{OSEP} = H^0_{KS} + |inlm\sigma\rangle\langle inlm\sigma|V_{ext} \quad , \tag{1}$$

where $H^0_{KS}$ is the original Kohn-Sham Hamiltonian which includes all the orbital-independent potential. The new secular equation with this new Hamiltonian then can solved the in the framework of DFT in a self-consistent way, without additional efforts. In such scheme multiple orbital-dependent potentials can be applied to the system simultaneously, providing great flexibility to study various effects on the physical or chemical properties of the system. Using such approach, Du et al. [150] have explained the formation of the lone pairs, which is very important for the ferroelectric systems involving Pb or Bi ions. In the following part, we will further demonstrate the power of the OSEP method by uncovering the origin of ferroelectricity in perovskite ferroelectric oxides.

## 3 First-principles study of magnetoelectric and multiferroic materials

In recent years, with the amazing development of information industry, as well as the pursuit of high-speed, high-density and low-energy-cost storage, multiferroics has attracted large number of interest in theoretical and experimental research due to its inspiring potential application of in data storage. Meanwhile, the multiferroics have become one of the hottest topics in condensed matter physics and materials science. In the following part, we will show some representative cases of the applications of first-principles calculations in the study of magnetoelectric and multiferroic materials.

### 3.1 First-principles investigations on single phase multiferroics

As pointed out before, the so-called multiferroic [2] material is one that possesses two or more kinds of the primary ferroic properties, i.e., ferroelectricity, ferromagnetism, ferroelasticity, and ferrotoroidicity in the same phase. Due to a great deal of attentions in the ordering of ferroelectricity and ferromagnetism, i.e., magnetoelectric effect, multiferroics are regarded as a special kind of magnetoelectric materials. Except for the inherent properties that a spontaneous magnetization (electrical polarization) can be reoriented by an applied magnetic (electric) field, multiferroic materials, where these two order parameters are intimately coupled, have the prospect of controlling charges by applied magnet-ic fields and spins by applied voltages [5]. Traditionally, such materials have all the potential applications of both their parent ferroelectric and ferromagnetic materials, such as transducers, actuators, and capacitors based on ferroelectricity, and sensors, read heads, spin transistors and magnetic valves based on ferromagnetism [151]. In addition, it is precisely because the two polarization vectors, i.e., magnetization and polarization, which are widely applied in encoding information, are closely connected in a single multiferroic material, the application of multiferroics in information storage can be envisaged. The concept of multi-state memory [152-154], multiferroic random access memory [155-157], and electrically assisted magnetic recording (EAMR) [158] has been pointed out and some of them have already been experimentally demonstrated. Based on the multiferroics, especially the coupling between magnetization and polarization, the functional devices and storage technologies possess tremendous potential applications in information industry.

Despite the attracting prospect and large amounts of research in this field, very few multiferroics exist in nature or have been synthesized in the laboratory. Furthermore, the weak coupling between ferromagnetism and ferroelectricity of the majority proved multiferroics, as well as the Curie temperature far away from room temperature, restrict the practical applications of multiferroics. Here we present several representative theoretical work to track the advances in the research of single phase multiferroics.

### 3.1.1 Origin of ferroelectricity in perovskite oxides

In some materials such as perovskite oxides ($ABO_3$), their nonzero misalign of positive and negative charge centers induces spontaneous electric dipole moments that can be reversed by the application of an external electric field. This property of spontaneous electric polarization is called ferroelectricity. The study of ferroelectric materials began in 1894 when anomalously large piezoelectric constants was observed in Rochelle salt. However, it was not until 1940s that ferroelectricity was observed in BTO and related perovskite-structure oxides after the report of phenomenological theories for dielectric, piezoelectric, elastic, and ferroelectric phase transition behaviors [159]. Cochran [160] and Anderson [161] proposed ferroelectric soft mode theory, which correlates phase transitions and lattice dynamical instability and has been proved in strontium titanate by using infrared reflection by Barker et al [162] at the University of California. The soft mode theory is quite motivational though we have known that the transition is weakly first order in many perovskite ferroelectrics such as orthorhombic $β$-phase BFO, which may be the consequence of polarization-strain coupling [163, 164].

Owning to the sensitivity to chemistry, defects, electrical boundary conditions and pressure, perovskite ferroelectrics display very different ferroelectric behavior, which results from a delicate balance between long-range Coulomb (Madelung) force (favors the ferroelectric state) and short-range repulsions induced by overlapping of electron cloud (favors the nonpolar cubic structure). Using first-principles approach of linearized augmented plane-wave method (LAPW) [165] within the LDA for the exchange and correlation interactions [166], Cohen et al. [167, 168] proposed an origin for ferroelectricity in perovskite oxides that hybridization between the B cation and O is essential to weaken the short-range repulsions and allows the ferroelectric transition. In most ferroelectric oxide perovskites containing B cations, the lowest unoccupied states are $d$-states, by hybridization with the O states, they consequently soften the B-O repulsion and allow the possibility of ferroelectric instability.

To shed light on the origin of ferroelectricity in perovskite oxides unambiguously, the above described OSEP method is adopted to investigate the ferroelectric instability in two typical perovskite oxides, i.e., BTO and PTO, by shifting the energy level of specific atomic orbital. As the strength of hybridization between two atomic orbits is strongly dependent on their energy difference, by applying an external field to shift the energy levels of these orbitals we can effectively weaken (or strengthen) the hybridization [150]. Therefore we can obtain the immediate consequence caused by change of the orbital hybridization.

As depicted in partial density of states of BTO in Fig. 1a, Ti $3d$ orbitals mainly dominates around 3 eV at the conduction band, and the remaining Ti $3d$ orbitals primarily locate at the valence band, hybridizing with O $2p$ orbitals to induce ferroelectricity. By exerting external field to Ti $3d$ orbitals we can easily control the hybridization between Ti $3d$ and O $2p$ orbitals. To address the issue intuitively, a double-well potential profile is adopted to describe paraelectric-ferroelectric phase transition. As shown in Fig. 1b, shifting up Ti $3d$ orbitals lowers the double-well potential, thus weakens ferroelectricity. In particular, when shifting the on-site orbital energy by 2 eV, the double-well potential vanishes, instead a single well potential crops up which indicates the disappearance of ferroelectricity. Therefore, the ferroelectricity of BTO is dependent on the hybridization between Ti $3d$ and O $2p$ orbitals, providing straightforward numerical support of Cohen's theory [168].

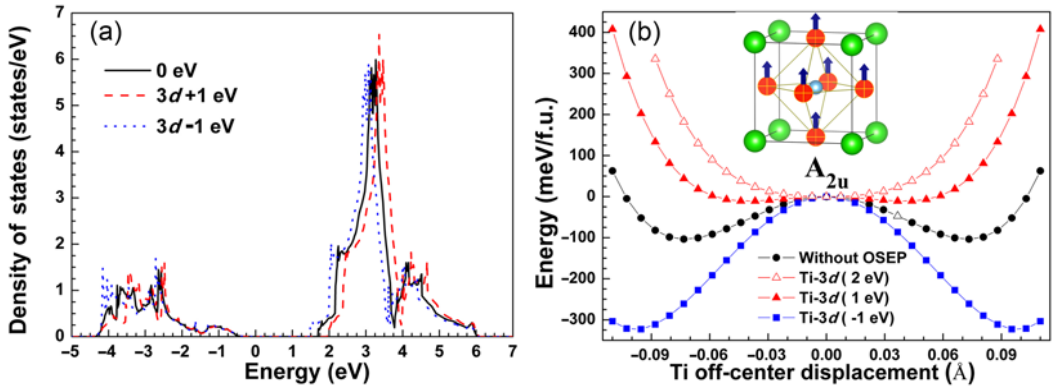

**Fig. 1** (Color online) **a** Density of states of BTO projected onto the Ti $3d$ orbitals with OSEP method, and the zero is set to the Fermi level. **b** Double-well potential of BTO with total energy as a function of Ti off-center displacement for lattice distortion ($A_{2u}$ mode), shifting Ti $3d$ states by the OSEP method. The inset shows corresponding soft mode

Another conventional ferroelectric is PTO, in which ferroelectricity is thought to be mainly induced by Pb $6s$ lone pair and the hybridization between Ti $3d$ and O $2p$ orbitals. The double-well potential obtained from general DFT method and OSEP-based DFT method by shifting Pb $6s$ and Ti $3d$ orbitals is shown in Fig. 2. As we can see, shifting down Ti $3d$ or shifting up Pb $6s$ orbital will increases the depth of double-well potential and enhances ferroelectricity, yet the influence of Pb $6s$ orbital is much less than that of Ti $3d$ orbitals. Therefore, both Pb $6s$ and Ti $3d$ orbitals play important roles in the ferroelectricity of PTO, which is also consistent with Cohen's theory [168].

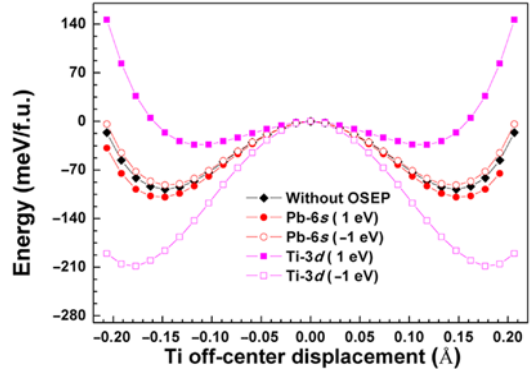

**Fig. 2** (Color online) Double-well potential of PTO with total energy as a function of Ti off-center displacement for lattice distortion, shifting Pb-6$s$ and Ti-3$d$ by OSEP method

### 3.1.2 The root of multiferroics for ABO$_3$ structures

In prototypical ferroelectric materials such as BTO and PTO, which both contain transition metal cations with a formal $d^0$ electronic structure, the requirement for "$d^0$-ness" has been explained in terms of covalent bond formation between empty transition-metal and filled O $2p$ orbitals. But magnetic system has partial occupation of the transition-metal $d$ orbitals. Therefore, magnetism and conventional ferroelectricity are believed mutually exclusive [151], at least in perovskite oxides. It is generally assumed that there are very few multiferroic materials which are both magnetic and ferroelectric. The contradiction between B-site ferroelectricity and B-site magnetism therefore demands new mechanisms for ferroelectricity in multiferroics, for example lone pair stereochemical activity, geometric ferroelectricity, spin spirals, etc. [3]. It is interesting to point out that recently Wan et al. [169] proposed an entirely different theory stating that short range magnetic interaction supports ferroelectric instability in magnetic oxides, which was exotic yet needs to be verified experimentally.

The ions $Bi^{3+}$ and $Pb^{2+}$ have $(6s)^2(6p)^0$ electronic configuration, which belongs to the lone pairs. The lone pair state is unstable and will give rise to a mixing between the $(6s)^2(6p)^0$ ground state and a low-lying $(6s)^1(6p)^1$ excited state, which eventually leads these ions to break the inversion symmetry. The lone pair stereochemical activity which requires a $p^0$ configuration is an example of the second-order Jahn-Teller effect, which is similar with $d^0$ configuration. From real-space visualization of electronic structure as shown in Fig. S1, Hill and co-workers [170] confirms that highly distorted structure is induced by lone pairs. Therefore, the ions with lone-pair electrons locate at A-sites in an ABO$_3$ perovskite structure, and the magnetic ions locate at B-sites, which can avoid the incompatibility between magnetism and ferroelectricity. The typical representatives of lone pair stereochemical activity are BFO and BiMnO$_3$ (BMO). Note that with the help of the OSEP method, we can more clearly identify the role of lone pairs in the ferroelectricity, as did in Ref. [150]. This study is still in progress.

Another famous multiferroic material is YMnO$_3$ which is both ferroelectric and A-type antiferromagnetic [37]. YMnO$_3$ is an anomalous ferroelectric oxide, where rehybridization and covalency play minor roles. Instead, long-range dipole-dipole interactions and oxygen rotations both cooperate to drive the system towards the stable ferroelectric state.

In fact a structural instability in YMnO$_3$ is mainly generated by size effects and geometric constraints. The space-filling and atomic coordination in the symmetry structure is not optimal, but can be improved by a small distortion that eventually breaks inversion symmetry. DFT calculations showed that the ferroelectric structure of YMnO$_3$ arises from an interaction between a polar Γ-point mode and a non-polar Brillouin zone-boundary mode that leads to a unit cell tripling. In addition, phonon frequencies and group theoretical analysis show that YMnO$_3$ belongs to an improper ferroelectric, where the K$_3$ phonon is strongly unstable and interacts with the Γ phonon leading to the observed polarization [171].

### 3.1.3 BiFeO$_3$

Among multiferroic materials, BFO with high transition temperatures well above room temperature ($T_C$ ~ 1100 K, $T_N$ ~ 640 K), owning to abundant physics deriving from the interaction between charge [172], spin [157], orbital [173] and lattice [174], has been probably the most studied multiferroic material to date [175]. The ground-state structure of BFO is rhombohedrally distorted with space group $R3c$ ($a$=5.63 Å, $α$=59.4°) [176] as shown in Fig. S2a–c, arising from the cubic perovskite structure by two successive distortions [177]: (i) counterrotations of adjacent oxygen octahedra along [111] direction and (ii) relative displacements of Bi, Fe and O atoms along [111]. The ferroelectric polarization in BFO is 90 μC/cm$^2$ with a large displacement of the Bi ions relative to the FeO$_6$ octahedra along [111] direction [38], and is much larger than conventional ferroelectricity. With the influence of epitaxial strain, BFO shows T-like phase, which has gigantic ferroelectric polarization (135 μC/cm$^2$) [178]. Bulk BFO has G-type antiferromagnetic ordering (Fig. S2d) due to the superexchange interaction, and both the inter- and intra-plane couplings are antiferromagnetic. However, considering the Dzyaloshinskii-Moriya (DM) interaction BFO has spiral magnetic ordering (Fig. S2d) [177], and can be induced to produce weak ferromagnetism by breaking the spiral magnetic ordering. The ferromagnetic plane is perpendicular to the ferroelectric polarization direction and the easy magnetization axis is coupled to ferroelectric polarization direction in BFO, which can be applied to electric field control of magnetism [179].

The perovskite BFO was first prepared in the late 1950s, but limitations in the quality of bulk crystals had hindered us from observing true polarization values. The current interest in BFO was caused primarily by a 2003 paper from Ramesh's group [74]. In high quality epitaxial BFO thin film prepared by pulse laser deposition technique, they found unexpectedly large remnant polarization and a substantial magnetization which is generally believed to be related to extrinsic effects such as defects or small amounts of impurity phases.

Using first-principles calculations Ederer and Spaldin [79] found that the large electric polarization of BFO was actually intrinsic in bulk BFO and was insensitive to epitaxial strain in stark contrast to many other conventional ferroelectricity. A systematic comparison of the strain dependence in various ferroelectrics, including BFO both in an $R3c$ phase and a hypothetical tetragonal phase with $P4mm$ symmetry (Fig. S3). It was performed that the effect of epitaxial strain for all investigated systems can be understood in terms of piezoelectric tensor and Poisson ratio. We should point out that our recent study, which has not been published yet, demonstrates that the Fe–O interactions have much important roles in the ferroelectricity of BFO.

Domains have a wide range of practical applications [180], including nonvolatile random access memories and pi-ezoelectric devices. In rhombohedral BFO the ferroelectric polarization can point along any of the four diagonals of the perovskite unit cell hence BFO has three different domain structures (71, 109 and 180°). DFT can successfully calculate the characteristics of ferroelectric domains. Recently Ren et al. [181] investigated single domain and multidomain configuration of BFO as shown in Fig. 3a and found that the calculated wall energies follow the sequence 0.053, 0.071 and 0.156 J/m$^2$ for the 109, 180 and 71° domain walls, respectively, particularly the 109° domain wall is found to be the energetically most stable domain. Under compressive strain the domain structures display phase transition between rhombohedral-like (R-like) and T-like BFO phases, (Fig. 3b), and different domain walls have different strain conditions. In the certain strain regions the multidomain structures adopt an R-like phase, while the single domain configuration has already transformed into its T-like phase.

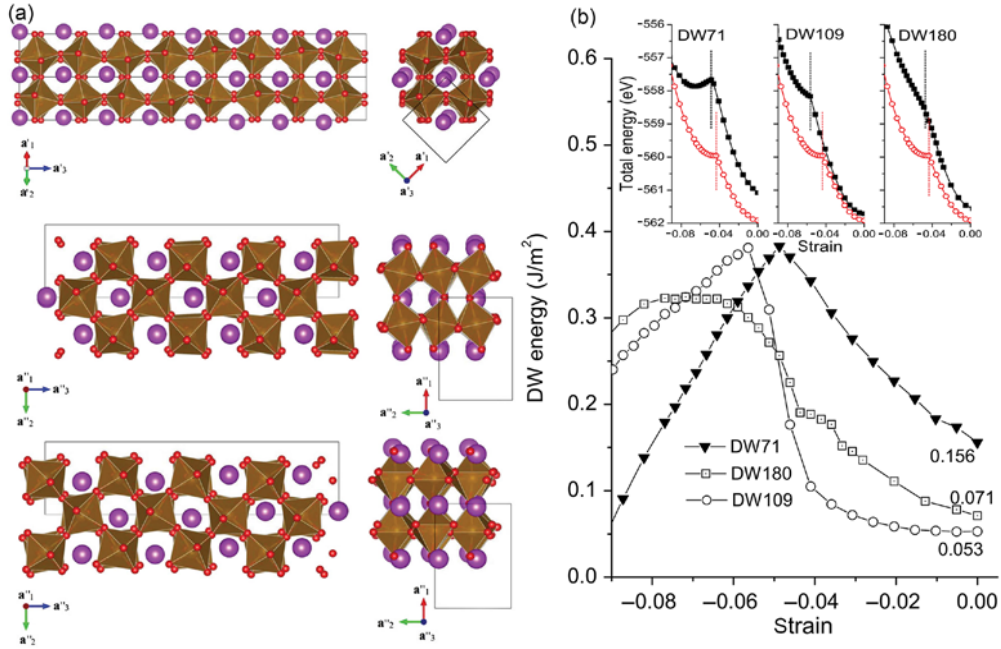

**Fig. 3** (Color online) **a** Schematic models for 109°(top), 71°(middle) and 180°(bottom) single domain and multidomain structures. **b** The relationship between domain wall energies and in-plane epitaxial strain for the investigated multidomain structures. The insets show the minimum total energy of the multidomain structures (in solid squares) and single domain configurations (via open circles) as a function of the misfit strain, and the vertical dashed lines represent the phase transition between R-like and T-like phase. Reproduced with permission from [181]. Copyright 2013, American Physical Society

Ding and Duan [35] studied the relationship between the magnetic ordering and ferroelectric polarization in the bismuth multiferroic oxides by theoretical calculations, and found a solution to the electric field control of the magnetic ordering of the T-like BFO. There exists a transition from C-type to G-type antiferromagnetic phase of T-like BFO with the in-plane constant 3.91 Å when the ferroelectric polarization is along the [001] direction. The magnetic ordering of T-like BFO is determined by the delicate balance between the Heisenberg exchange constant $J_{1c}$ and $J_{2c}$, which could be influenced by the ferroelectric polarization. Fig. 4 illustrates the evolution of the magnetic ordering with the change of the lattice constant and the ferroelectric polarization. In the shaded zone of Fig. 4, ferroelectric polarization change transforms C-type to G-type antiferromagnetic phase. The properties of spin transport of C-type and G-type antiferromagnetic state are entirely different along $z$ axis, which makes it possible to modulate the magnetic resistance of a single phase multiferroic film to record information by electric field.

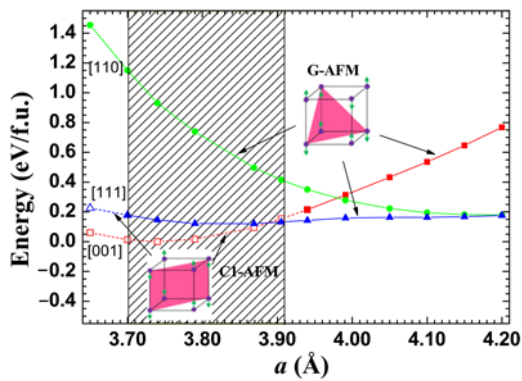

**Fig. 4** (Color online) Total energies, respect to the energy of [001] polarization at 3.74 Å, of the T-like BFO with ferroelectric polarization along [001] (square), [111] (triangle), and [110] (circle) directions. The straight lines with solid symbols and dotted lines with open symbols indicate the magnetic orderings are G-type and C1-type antiferromagnetic, whose spin arrangements are shown as insets, respectively. Reproduced with permission from [35]. Copyright 2012, Institute of Physics

First-principles calculations are helpful to elucidate the interface structures of BFO related heterostructures. Huang et al. [104] investigated coherent super-tetragonal BFO/LaAlO$_3$ and rhombohedral BFO/LaAlO$_3$ heterointerfaces by transmission electron microscope and first-principles calculations. In Fig. 5a, there are three transition layers for BFO from cubic-like to super-T-like, a continuous expansion of the BFO crystal lattice, elongation of oxygen octahedra, and displacement of Fe ions at heterointerfaces. A similar pinned interface transition layer about two unit cells thick is directly observed at heterointerfaces. As shown in Fig. 5b, T-like BFO experiences compressive strain, and R-like BFO experiences tensile strain. Therefore, it exists a critical mismatch around –4.3 % relative to R-BFO. The phase stability of R-like and T-like BFO is a result of competition between the tilting of FeO$_6$ octahedra and the displacement of Fe ions, which is depended on the strain. These results are consistent with first-principles calculations.

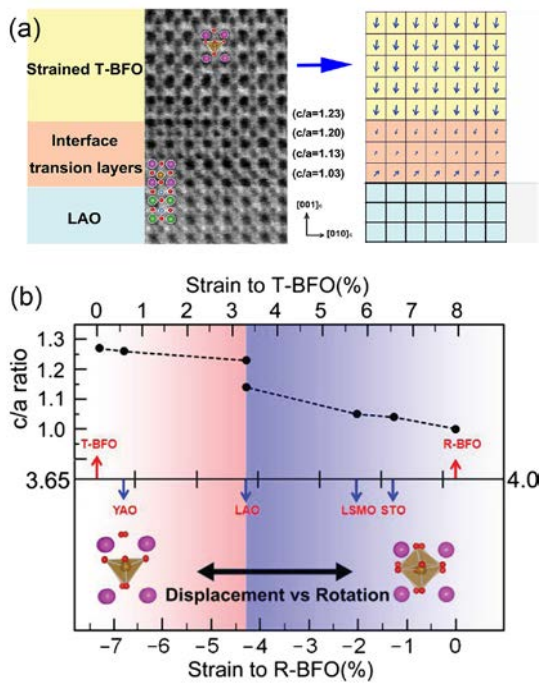

**Fig. 5** (Color online) **a** Transmission electron microscope image and polarization pinning diagrams for BFO/LAO interfaces and the corresponding polarization relaxation. **b** Substrate strain effect on the stability of T-BFO and R-BFO phases. Reproduced with permission from [104]. Copyright 2013, Wiely

### 3.1.4 Prediction of novel multiferroics

First-principles study, as discussed in previous section, is powerful to predict novel multiferroics. Here we introduce several examples. Some of them are already confirmed by recent experiments.

Bulk phase $EuTiO_3$ is an antiferromagnetic-ferroelectric with cubic perovskite structure. But in 2006, using first-principles density-functional theory calculations, Fennie and Rabe [182] reported that the transition from antiferromagnetic-paraelectric to ferromagnetic-ferroelectric phase will occur for biaxial compressive strain above 1.25 % (Fig. S4). As the value of the strain is easily attainable in experiments, the system targeted, thin-film $EuTiO_3$ strained in biaxial tension at +1.1 % on the $DyScO_3$ substrate, was experimentally proved to exhibit strong ferromagnetism (spontaneous magnetization, ~7 $\mu_B$ per Eu) and strong ferroelectricity (spontaneous polarization, ~29 ± 2 μC/cm) simultaneously in 2010 [183]. As we can see, first-principles study demonstrates an efficient strategy to find multiferroics under constrained circumstances.

Similarly in 2010, Bousquet et al. [19] theoretically predicted highly epitaxial strained ferromagnetic EuO to become ferroelectric and pointed out that its magnetic state remains ferromagnetic through the ferroelectric region. As the paramagnetic-to-ferromagnetic transition temperature of EuO is approximately 69 K, which is greater than that of $EuTiO_3$, i.e., (4.24 ± 0.02) K and will be further increased under epitaxial strain [184], EuO could be a potential candidate for multiferroics. Unfortunately, the calculated critical epitaxial strains from first-principles study for EuO, i.e., –3.3 % biaxial compression and +4.2 % biaxial tension (See Fig. S5), are too large to be realized experimentally at present.

Different from the prediction of strain induced multiferroic in EuO and $EuTiO_3$, octahedron rotations induced multiferroicity in Ruddlesden–Popper material $Ca_3Mn_2O_7$ (Fig. S6a) was predicted by Benedek and Fennie [185]. They found that two different octahedron rotation modes could drive the ferroelectric distortion and there exited a direct indication of improper coupling between the polarization, rotations, and tilts. In $Ca_3Mn_2O_7$ system, the two rotation modes with different symmetries were found to be the oxygen octahedron rotation mode with irreducible representation (irrep) $X_2^+$ and the oxygen octahedron tilt mode with irrep $X_3^-$ (Fig. S6b, c). The results from first-principles calculation show that the polarization becomes non-zero only when both rotations condense, indicating that the polar state of the system is driven by the combination of the two modes [186]. Also it is observed that the applied strain can influence this special mechanism of the ferroelectricity. Note that the octahedron rotation modes in $Ca_3Mn_2O_7$ induce not only the ferroelectricity, but also the weak ferromagnetism, it is very interesting to find some single phase multiferroic materials under this mechanism.

### 3.1.5 Design of artificial multiferroics

Not only can first-principles study predict unknown multiferroics, but also it helps to design multiferroics for us. For the lack of multiferroics in nature, many attentions are focused to accomplish multiferroics in artificial materials in latest years. As stated above, first-principles study precisely predicted strain induced multiferroicity in EuTiO$_3$, which paves a new road to design multiferroics by strain engineering. In fact before this breakthrough in multiferroics, biaxial strain due to the lattice mismatch between a thin film and the underlying substrate had extensively been used to tailor the physical properties of materials, such as altering the paraelectric-to-ferroelectric or paramagnetic-to-ferromagnetic transition temperature $T_c$ by hundreds of degrees, and improve the mobility of semiconductors in transistors [187].

In addition, a more active way to improve both magnetic and ferroelectric properties is doping [188]. A lot of work on the doped BFO has been presented in the past decade years. For B-site substitutions the Fe$^{3+}$ ions are replacing by other transition metal ions. A good example of such a case is a double perovskite of the Bi$_2$FeCrO$_6$ [189]. Generally, based on Kanamori rules, the superexchange interaction of Fe–O–Cr can induce ferrimagnetic ordering [190, 191]. Nevertheless, Bi$_2$FeCrO$_6$ shows a ferrimagnetic state at the ground state since the Fe–O–Cr bond angle deviates from the perfect 180 ° to satisfy Kanamori rules. The magnetic moments of Fe and Cr ions are quite different (5 $\mu_B$ for Fe$^{3+}$ and 3 $\mu_B$ for Cr$^{3+}$), it could induce a net magnetic ordering of 2 $\mu_B$ per Fe-Cr pair. The space group of Bi$_2$FeCrO$_6$ is reduced to $R3$ due to the breaking of compositional symmetry, and the ferroelectric polarization is about 90 μC/cm$^2$.

In the doping system an interesting artificial structure can be built, for example checkerboard model. Ferroelectric antiferromagnetic bulk BFO and ferromagnetic bulk BMO are good candidates for a nanocomposite with multiferroic behavior, i.e., ferroelectric and ferrimagnetic. Pálová et al. [192] designed BFO/BMO checkerboard model (See Fig. S7) and reduced the energy difference between its antiferromagnetic and ferromagnetic state, making it much more plausible for a magnetic transition to occur. The ferroelectricity can be preserved. However, such structure is hard to grow experimentally.

3.1.6 General theory for the ferroelectricity in spin-spiral multiferroics

In last ten years a serious of manganese oxides (RMnO$_3$ [193-196] and RMn$_2$O$_5$ [197, 198], with R = Y, Tb, Dy, Lu, etc.) have turned out to be multiferroics unexpectedly, leading manganese oxide to be an active field in multiferroics research. Employing first-principles calculations, theoretical researchers contributed many efforts to the exploration of single-phase multiferroics. Before 2007, It is a hard and time-consuming work to make a direct comparison of physical quantities for TbMn$_2$O$_5$ (TMO) between computational investigations and experiments because of the complicated incommensurate antiferromagnetic (AFM) structure with the propagation vector K = (0.48, 0, 0.32), but Wang et al.'s [199] work opened up a new way by constructing a relative simple structure with approximately equivalent propagation vector $K$ = (0.5, 0, 0) to identify the origin of ferroelectricity in two degenerate TMO (denoted by L and R in their publication). Fig. 6 is the energy surfaces versus displacements, which clearly depicts the physical picture of coupling between magnetic order and lattice. The high-symmetry structure H in Fig. 6 is an unstable point on the energy surface, hence it will tend to be L or R spontaneously according to its spin configuration of high-symmetry point. It's clear that even if without invoking the spin-orbit coupling (SOC) and noncollinear spins, the ferroelectricity can be fundamentally driven by the noncentrosymmetric magnetic ordering in TMO.

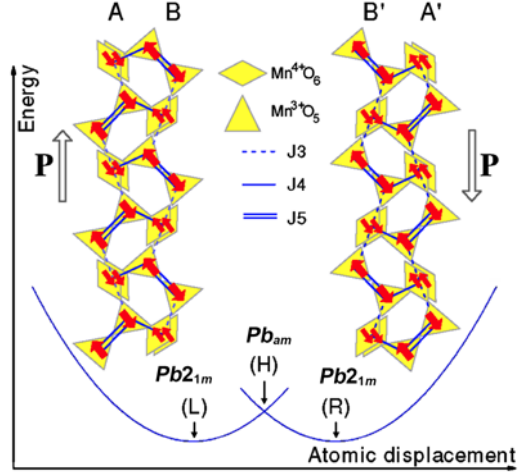

**Fig. 6** (Color online) An asymmetric-spin-chain model showing the sketch of energy surfaces of structure L and R vs the atomic displacements from the high-symmetry structure. The diamonds and triangles denote the $Mn^{4+}O_6$ octahedra and $Mn^{3+}O_5$ pyramids, respectively. Reproduced with permission from [199]. Copyright 2007, American Physical Society

Besides ferroelectricity induced by the noncentrosymmetric collinear magnetic order, it also exists in many other insulating helimagnets or spiral-spin structures. Recent years, DM interaction [200] or equivalent Katsura, Nagaosa, and Balatsky (KNB) spin-current model [201] has provided successful explanations for spiral-spin ferroelectrics. Within this model, $P$ is perpendicular to the direction of the magnetic modulation $q \propto e_{ij}$ (i.e., $P \perp q$). However, the experimental studies of multiferroics $CuFeO_2$ and $MnI_2$ shows abnormal results that $P$ in the helical spin-spiral state with $q = (Q, Q, 0)$ is parallel to $q$ (i.e., $P \| q$), which can hardly be interpreted by the symmetric exchange striction or spin-current model. To resolve the aforementioned question, Xiang et al. [202] developed a generalized KNB (gKNB) model for the ferroelectric polarization induced by spin spiral on the basis of symmetry considerations and verified its validity by studying on $MnI_2$ with first-principles calculation. As shown in Fig. 7a, the DFT calculations shows that $P \perp q$ alone the [100] direction and $P \| q$ alone [110] direction, with $q = (1/3, 0, 0)$ and $q = (1/3, 1/3, 0)$, respectively, which corresponds to Kurumaji's experiment [203]. Also, as shown in Fig. 7c, d, the gKNB model succeeds in predicting the correct direction of the electric polarization and giving a rather accurate magnitude of the polarization for the cases of $q = (Q, 0, 0)$ and $q = (Q, Q, 0)$, which manifests its applicability for ferroelectricity driven by spin-spiral magnetic order.

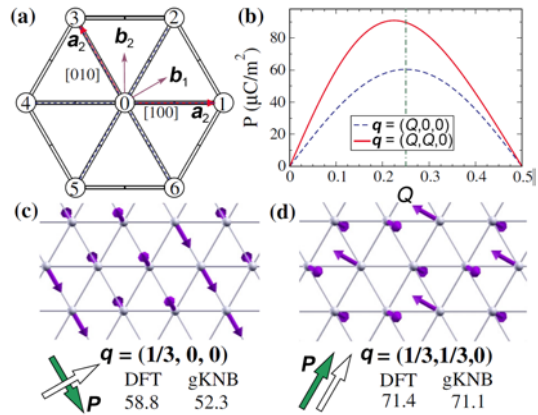

**Fig. 7** (Color online) **a** The triangular lattice of $Mn^{2+}$ ions, where the in-plane lattice vectors $a_1$ and $a_2$ and the corresponding reciprocal lattice vectors $b_1$

and $b_2$ are shown. **b** The magnitude of the polarization predicted from gKNB model as a function of $Q$ for the cases of $\boldsymbol{q} = (Q, 0, 0)$ and $\boldsymbol{q} = (Q, Q, 0)$. **c, d** The spin orientations of two proper-screw spirals with $\boldsymbol{q} = (1/3, 0, 0)$ and $\boldsymbol{q} = (1/3, 1/3, 0)$. The modulation vector $\boldsymbol{q}$ and the polarization vector $\boldsymbol{P}$ are represented by the white and green arrows, respectively. The numbers (in $\mu C/m^2$) denote the magnitudes of the polarizations obtained from the direct DFT calculation and gKNB model. Reproduced with permission from [202]. Copyright 2011, American Physical Society

Furthermore, Xiang et al. [204] extended gKNB model to noncentrosymmetric spin dimers and investigated the origin of the giant ferroelectric polarization of $CaMn_7O_{12}$ using DFT calculations. Different from the ferroelectricity in other multiferroic systems such as $LiCu_2O_2$ [205] purely caused by the SOC effect, the origin of ferroelectricity of $CaMn_7O_{12}$ involves more fundamental physics. In $CaMn_7O_{12}$, the magnitude of the ferroelectric polarization is determined by the exchange striction, but the sign is basically controlled by the strong DM interaction. Consequently, the giant ferroelectricity in $CaMn_7O_{12}$ is driven by strong DM interaction, combining with exchange striction.

### 3.2 Exploring novel magnetoelectric effect

In order to accomplish the electric control of magnetism, one major scheme is to explore the heterostructures with ferroelectric and ferromagnetic compounds [206-209]. In these structures the applied electric field can induce a strain in the ferroelectric layers. The strain can be transferred into the adjacent ferromagnetic layers through the magnetostrictive coupling, and further influence the magnetism in the ferromagnetic compounds [6, 210]. Novel magnetoelectric effect, however, could occur when the ferroelectric/ferromagnetic interfaces are epitaxial. This is firstly pointed out by the first-principles calculations, and will be discussed as followed.

### 3.2.1 Interface magnetoelectric effect

Recently, it has been found that the reversal of polarization in the ferroelectric layers can also affect the magnetic properties at the interface of ferroelectric/ferromagnetic compounds, which is called the interface magnetoelectric effect (ME) [211, 212]. In 2006, Duan et al. [110] studied the heterostructure of Fe/BTO (Fig. S8a). In this structure, when the BTO is in the paraelectric state, the magnetic moments of the interfacial atoms are exactly the same at the bottom and top interfaces due to the symmetry. However, in the ferroelectric state, the magnetic moments are different between the interfacial atoms at the bottom interfaces and those at the top ones. For Fe atoms, the one at the bottom interface has an enhanced magnetic moment, while the magnetic moment at the top interface is reduced. Similarly, the magnetic moments of the Ti atoms at the two interfaces produce a sizeable asymmetry. By analyzing the orbital-resolved DOS for the interfacial atoms, they found that there exists a strong hybridization between Fe in minority spin states and Ti $3d$ orbitals, which induce a magnetic moment of Ti atoms. When the BTO is polarized upward, the Ti atom moves toward the top interface and enhances the bonding between Fe and Ti $3d$ orbitals at the top interface, which lead to the minority-spin bonding state lies in deeper energy level. Such effect can directly increase the induced magnetic moment of top Ti atom and decrease the magnetic moment of top Fe atom. Fig. S8b shows the minority-spin charge density by switching the polarization, which can reflect the dissimilar behaviors of magnetic moment at the two interfaces. As seen from the Fig. S8b, it is clear that the charge distribution has a much higher overlap between the Fe atom and Ti atom for the up polarization than the down polarization. Considering the change of the interface magnetic moment, they obtained the magnetoelectric coefficient $\alpha = \mu_0 \Delta M / E_c \approx 0.01$ G cm/V, which is of the same order in magnitude as epitaxial BFO/CoFe$_2$O$_4$ columnar nanostructures [213].

Besides adjusting the interface magnetic moment, another interesting phenomena is to utilize this interface ME to manipulate the MAE of the ferromagnetic metal films. In 2008, Duan et al. [214] also focused on Fe/BTO heterostructure, and found that the polarization in BTO can influence the orbital magnetic moments and MAE of the Fe layers at the interface. As shown in Fig. S9, when the polarization is reversed, the MAE changes as much as 50 %. The main reason of this change in MAE also comes from the displacement of the atoms at the interface under the polarization, which can increase the hybridization between the Fe and Ti minority-spin 3*d* orbital and thus modify the electron distribution in these orbitals. Following Bruno [215] and van der Laan's [216] model, they obtain a similar result (Fig. S9), proving the rationality of the first-principles calculation. After that, motivated by the theoretical achievements, Sahoo et al. [217] investigated this interface ME in epitaxial thin film Fe/BTO. They found that magnetization changes greatly in the ferroelectric switching and structure transitions of BTO by electric field. Also there exists up to 20 % of change in coercivity field by different electric field, indicating that the interface magnetic anisotropy has been altered. It is assumed that this discovery can be applied to design films with the switchable magnetic anisotropy and is helpful in the area of reading and writing on media with high perpendicular uniaxial anisotropy constant.

3.2.2 Spin-dependent screening induced magnetoelectric effect

The interface ME mentioned above mainly comes from the displacement of the interface atoms under the electric polarization. In other heterostructures, some different mechanism of interface ME has been studied. In 2007, Rondinelli et al. [218] explored the interface ME in SrTiO$_3$/SrRuO$_3$/SrTiO$_3$ (STO/SRO/STO) by first-principles calculation. As shown in Fig. 8a, by the application of the electric field, the accumulation of spin is localized at the interfaces with an equal magnitude and opposite sign for two electrodes. The spin response of the interface is measured in terms of ratio of the surface spin polarization to the surface charge density $\eta$, which is about 0.37 for the STO/SRO/STO system. Fig. 8b shows the spin density distribution at the high frequency regime, in which case, the ions are fixed and structural effects on magnetism can be eliminated. Although the induced spin density and magnetic moment are 10 times less than the static case, the value of $\eta$ keeps constant, which indicates that this ME coupling is only ascribed to the accumulation of spin carriers at the interface and does not originate from the spin-lattice interaction. Also by replacing STO to BTO, the ratio $\eta$ is unchanged, demonstrating that the larger amount of charge to screen the ferroelectric polarization can induce a larger change of the interface magnetic moment. Note that the phenomena that spin-polarized charge is stored at the interface asymmetrically like the behavior of the charge in traditional capacitor. This study provides a new spin capacitor for the analogous spintronics devices in the future.

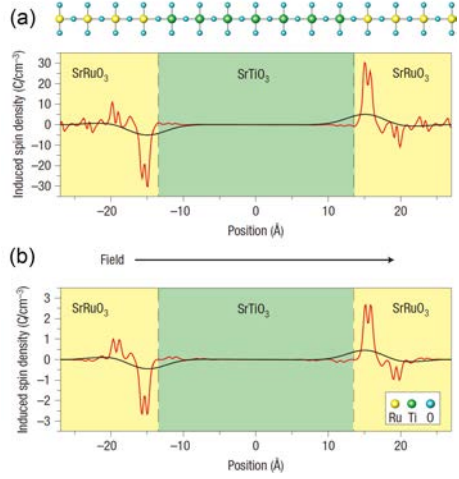

**Fig. 8** (Color online) Calculated magnetization induced by an external voltage in a nanocapacitor consisting of seven layers of STO alternating with seven layers of ferromagnetic, metallic SRO electrodes with parallel magnetic alignment. The static response including ionic and electronic contributions (**a**) and the high-frequency (electronic only) response (**b**) are shown. The smooth and zigzag curves show the planar averaged and macroscopically averaged induced magnetizations, respectively. Reproduced with permission from [218]. Copyright 2008, Nature Publishing Group

Besides exploring the interface ME in heterostructures, another major scheme is to apply the electric field directly in ferromagnetic metals [67, 219]. Considering the spin accumulation in ferromagnetic metals under the applied electric field is limited at the surface of the film due to the spin-dependent screening [220], this magnetoelectric effect is called surface ME. In 2008, Duan et al. [66] investigated the Fe(001), Ni(001) and Co(0001) thin film under the external electric field by first-principles calculation. They found that under the external electric field a change of the magnetic moment was induced at the surface layer, and the magnetic moment of the surface Fe atom displayed a linear relationship with the applied field. By setting up a model which assumes that the screening charge is localized at the metal surface and the free charge carriers at surfaces are spin-polarized, they obtained the surface ME coefficient $\alpha_s$ as the following description:

$$\alpha_s = \frac{\varepsilon \mu_B}{ec^2} \frac{n^\uparrow - n^\downarrow}{n^\uparrow + n^\downarrow}, \tag{2}$$

where $n^\uparrow$ and $n^\downarrow$ are the surface DOS of majority-spin and minority-spin states at the Fermi level and $\varepsilon$ is the dielectric constant adjacent to the ferromagnetic film. From this equation, it is clear that the surface ME coefficient is decided by the spin distribution of the surface DOS. Fig. S10 shows the induced spin charge density along the Fe films. It can be seen that there exists two opposite spin density at the two surfaces which lead to two different behavior in magnetic moment of surface Fe atoms under the electric field. Besides the change of magnetic moment, the surface orbital moment anisotropy and MAE increased linearly with the external field, which provide a new way to adjust the MAE electrically.

Above introductions mainly focus on the surface ME in the ferromagnetic films, where both the majority- and minority spin channels have the surface states at the Fermi level. Thus the ME coefficient is a fraction of the two surface DOS from Eq. (2). However, for half-metals, where only one spin channel is conducting, according to Eq. (2), the surface ME coefficient $\alpha_s$ can be written as

$$\alpha_s = \pm \frac{\mu_B}{ec^2} \approx \pm 6.44 \times 10^{-14} \frac{\text{G cm}^2}{\text{V}}, \qquad (3)$$

suggesting that the ME coefficient of half-metal is universal constant and has nothing to do with the specific electric, crystal and surface structure. In 2009, Duan et al. [221] studied the surface ME of the $CrO_2$ thin film, which maintains the typical half-metallicity at the (001) surface in the bulk. From the DOS of Cr in Fig. 9a, they found that only the majority-spin states exit at the Fermi level, and the minority-spin states exhibited an energy gap, indicating that $CrO_2$ was half-metal both in bulk and thin film structure. Fig. 9b shows the distribution of the spin density of majority- and minority-spin in the film under the external electric field. It is seen that the surface spin accumulation is mainly contributed by the majority-spin channel and the net induced minority-spin charge at each surface is nearly zero because of its half-metallicity. By calculating the linear slope of the change in magnetic moment with the applied electric field, they pointed out that the surface ME coefficient $\alpha_s$ is $-6.41 \times 10^{-14}$ G cm$^2$/V, which is identical to the prediction from Eq. (3). Because this unique property in surface ME may happen only in half-metals, it would be a new way to detect half-metallicity by measuring the surface ME coefficient in the future.

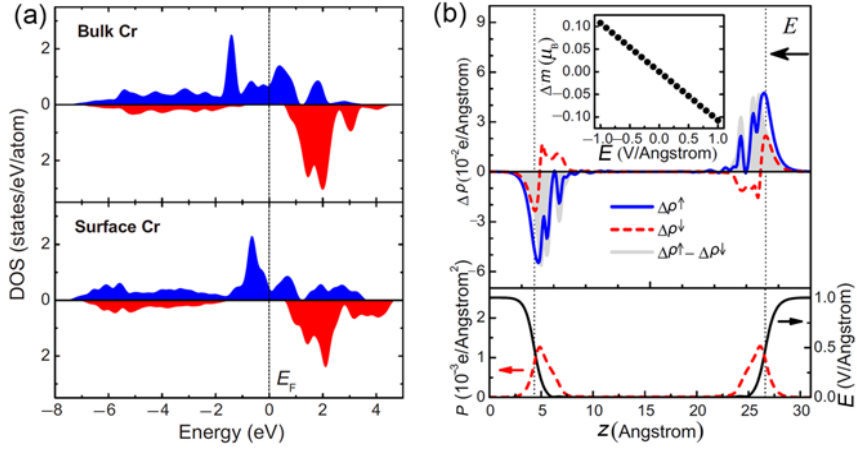

**Fig. 9** (Color online) **a** Spin-polarized DOS at the bulk and surface Cr atoms in a 15-monolayer $CrO_2$ (001) film; **b** top: induced spin-dependent charge densities $\Delta\rho = \rho(E) - \rho(0)$ for majority-(↑) and minority-(↓) spin electrons and spin density $\Delta\rho^\uparrow - \Delta\rho^\downarrow$ average over the plane. Down: variation in the electric field and the induced minority-spin polarization across the film. Inset: induced magnetic moment $\Delta m$ per unit cell area of the $CrO_2$ (001) surface as a function of the applied electric field. Reproduced with permission from [221]. Copyright 2009, American Physical Society

Note that the surface ME coefficient in the systems mentioned above only has a small value, in order to enhance it and accomplish the control of magnetism by a small applied electric field, some investigations have been carried out. From Eq. (2), one possible way to increase $\alpha_s$ is to improve the dielectric constant $\varepsilon$. In 2010, Niranjan et al. [222] performed first-principles calculation to explore the influence of the external electric field on magnetic properties of the Fe/MgO interface. By considering the linear slope of magnetic moment of Fe atom at the Fe/MgO, they found that ME coefficient is much larger than previous report in Fe (001) surface [66]. Also it is seen that the interface MAE changes with electric field and displays a more obvious slope of change comparing with the Fe (001) surface. Fig. S11 shows the induced spin density in the (010) plane along the Fe/MgO interface and the change of magnetic moment in the interface Fe atom under the external electric field. They assumed that the behavior in ME coefficient mainly comes from the fact that the electric field is applied across the MgO layer, which is a kind of typical dielectric materi-als with a high dielectric constant. By analyzing the value of ME coefficient, they found that the ME at the Fe/MgO interface is enhanced with the Fe (001) surface by a factor of 3.8, which is equal to the high frequency dielectric constant of MgO approximately. Such calculation indicates that the dielectric materials can increase the induced spin density at the interface under the electric field and further improve the ME at the interface. They also predicted that there existed a more significant influence of electric field on the interface MAE due to the change of the relative electron occupation in 3$d$ orbitals of the Fe atom at the Fe/MgO interface. Inspired by the theoretical prediction, Maruyama et al. [223] grew a thin bcc Fe(001)/MgO(001) junction and found that a great change in the hysteresis curve of magneto-optical Kerr ellipticity with the magnetic field happened under a small electric field (less than 100 mV/nm), indicating that the MAE is changed significantly. These experimental and theoretical results would be useful to develop the electrically written magnetic information technology in the future.

### 3.3 Ferroelectric and multiferroic tunnel junctions

Tunnel effect is a well-known phenomenon in quantum mechanics. It states that a particle, e.g. electron, has possibility to travel through a potential barrier that is higher than the particle's energy. This effect not only has fundamental scientific interests, i.e., it demonstrates wave-particle dualism, but is of great importance to modern technology. For instance, tunnel magnetoresistance (TMR) [224], or more specifically the magnetic tunnel junction (MTJ), is now widely applied in hard-disk drives and magneto resistive random-access memory (MRAM) [225]. The successful application of TMR has been inspiring people to study other types of tunnel effects. Considering the amazing analogy between ferromagnetism and ferroelectricity, it is natural to connect tunnel effect with ferroelectricity, and even with ferroelectricity and ferromagnetism together, i.e., ferroelectric tunnel junctions and multiferroic tunnel junctions.

### 3.3.1 Ferroelectric tunnel junctions

Early in 1970s, polar switch [226] involving a switchable thin-film ferroelectric material was proposed by Japanese physicist Esaki (Nobel Prize in Physics, 1973) in IBM. It was thought to be the earliest ferroelectric tunnel junction and expected to be an ideal candidate for nonvolatile memory cells. However, the research of polar switch in IBM was finally abandoned when Batra and Silverman [227] proved that the depolarization field would modify the thermodynamic properties of ferroelectrics and thin film ferroelectric materials less than 400 nanometers were thermodynamically instability.

The nanoscale ferroelectric films was not observed until Bune et al. [228] reported the ferroelectric transition in two dimensional crystalline films of a random copolymer of vinylidene fluoride and trifluoroethylene [P(VDF-TrFE)] just 10 Å (two monolayers) thick. This inspiring breakthrough of nanoscale ferroelectricity was further confirmed in many other kinds of ferroelectrics (PZT by Tybell et al. [229]; PTO by Rabe et al. [230]; BTO by Sai et al. [231]; STO by Son et al. [232]; BFO by Rault et al. [233]) experimentally and theoretically. At present, the observed size of ferroelectric films has declined down to 0.4 nm [234]. For convenience, we list some milestones of nanoscale ferroelectricity in history (Fig. 10).

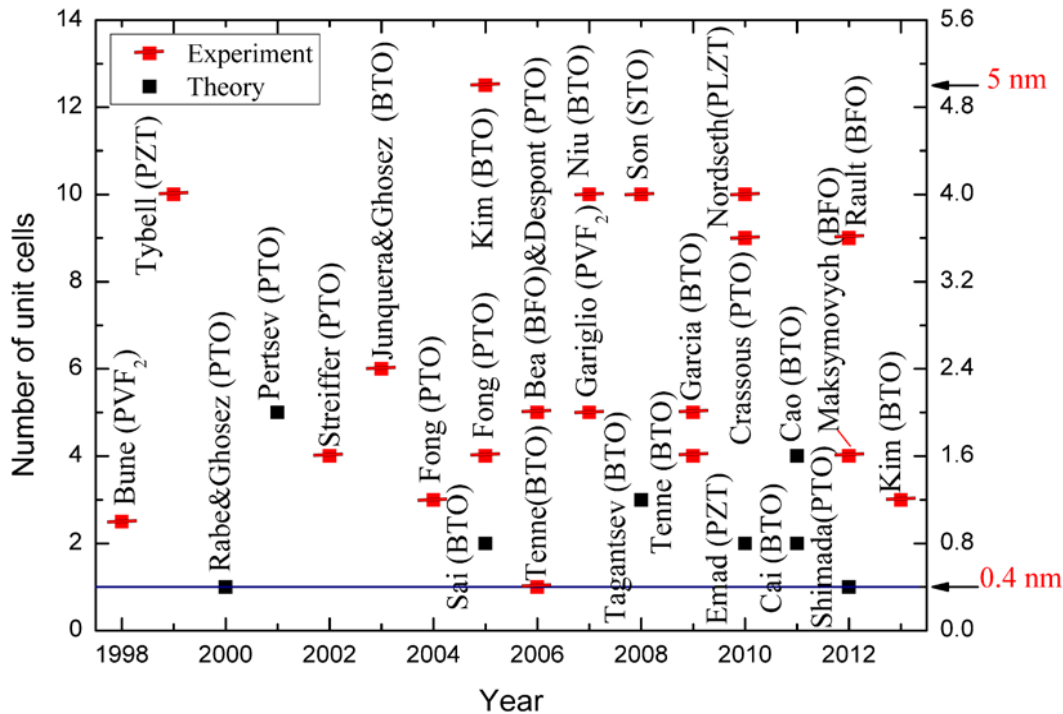

**Fig. 10** (Color online) Some milestones of the ferroelectricity at the nanoscale in the history

Benefiting from the notable evolution of the ferroelectricity at the nanoscale, the concept of ferroelectric tunnel junction was formally proposed by Kohlstedt et al. [235] in 2001 MRS Fall Meeting. In 2003, Contreras et al. [236] fabricated metal-ferroelectric-metal tunnel junctions with a $Pb(Zr_{0.52}Ti_{0.48})O_3$ film thickness ranging from 4 to 6 nm and reported two well-defined stable and reproducible resistance states in a 6-nm-thick $Pb(Zr_{0.52}Ti_{0.48})O_3$ thin film. They suggested that the polarization reversal in the ferroelectric barrier should be the origin of the observed resistive switching and the phonon-assisted inelastic tunneling processes were likely to be the dominant transport mechanism through 6-nm-thick $Pb(Zr_{0.52}Ti_{0.48})O_3$ thin film. What need alludes is, Contreras [237] wrote his doctoral thesis with the name of "Ferroelectric Tunnel Junctions" in the same year. This work presented a more detailed description of the concept of ferroelectric tunnel junctions and reviewed some significant experimental results and theoretical investigations of the quantum-mechanical electron tunneling through ferroelectric tunnel barriers.

Besides the importance of the nanoscale ferroelectricity, transport mechanism through the ferroelectricity is quite crucial for a ferroelectric tunnel junction. Tsymbal and Kohlstedt [238] thought polarization reversal in ferroelectric barriers exerts strong effect on the conduction of ferroelectric tunnel junctions. In a ferroelectric tunnel junction shown in Fig. S12a, polarization reversal alters the sigh of polarization charges at the barrier-electrode interface, which consequently changes the depolarization field and potential profile seen by the transport electrons because of incomplete screening (Fig. S12b). Using first-principles calculation, Duan et al. [109] demonstrated that interface effect could strongly influence the ferroelectricity of ferroelectric perovskite thin films. In fact, the interface effect also modifies the hybridization at the interface hence influences the electron transmission possibility (Fig. S12c). In addition, Kohlstedt's theoretical calculations showed that strain effect (Fig. S12d) brought by piezoelectricity under an applied voltage could change transport characteristics of the barrier such as the barrier width and the attenuation constant [239].

At present, first-principle study, combining with nonequilibrium Green's function's approach or phenomenological model has been thought to be a powerful tool to demonstrate the fundamental relations between transport, interface, electrodes' screen and polarization in barrier, etc. for ferroelectric tunnel junctions. Luo et al. [240] reported a tunable tunnel electroresistance (TER) effect by mechanical loads. They used the fully self-consistent nonequilibrium Green's function method combined with first-principles DFT calculation to investigate the TER induced by applied strain. They showed the applied strain not only could induce a stable and robust polarization, but also modify the effective potential barrier to accomplish paraelectric-ferroelectric phase transition. Therefore, an enhanced TER effect could be modulated by applied strain in their design. Hinsche et al. [241] calculated the TER for ferroelectric tunnel junction with BTO and PTO barriers by combining the microscopic electronic structure of the barrier material with a macroscopic model for the electrostatic potential. Their study clearly indicated the TER ratio is in dependence on the intrinsic polarization, the chemical potential, and the screening properties of the electrodes while the sign of the TER ratio depends on the chemical potential. What should be pointed out here is that their calculation method provide a flexible and reliable computational way to investigate the transport properties of ferroelectric tunnel junctions because their calculation can observe the influence of material-specific parameters, like the electrodes' properties, on the TER effect while taking the electronic structure of the barrier into consideration and does not need a whole electronic structure calculation for the whole junction. In addition, several good reviews written by Scott (2006) [242] and Fang et al. (2013) [243] are available for readers to find more information about the transport mechanisms in ferroelectric tunnel junctions.

### 3.3.2 Multiferroic tunnel junctions

For the purpose of realizing more functionality in ferroelectric tunnel junctions, the normal metal electrodes can be replaced by ferromagnetic materials. The interplay between the ferroelectric order, ferromagnetic order and ferroelastic order makes these spintronics devices be multiferroic tunnel junctions (MFTJs), which have been fabricated by many groups [152, 244, 245]. Based on first-principles calculations, Velev et al. [154] predicted four well-defined resistance states in SRO/BTO/SRO MFTJ (See Fig. S13). What is worth mentioning in their work is that the two asymmetrical interfaces in this MFTJ provides the crucial structure basis for the coexistence of TER and TMR. The theoretical result of four states in MFTJs is in agreement with the work of Gajek et al. [152] and has been observed in some other MFTJs [246-248]. These results reveal the exciting prospects of such systems for application in multilevel nonvolatile memories, tunable electric and magnetic field sensors, and multifunctional resistive switches.

Layer-by-layer method has been suggested as a new route to multiferroics and MFTJs, as it circumvents the contra-indication between magnetism and ferroelectricity by constraining the magnetic ions in a polar arrangement in spite of their natural tendency to remain centrosymmetric [4]. In this way, we have more alternative materials to act as barriers in MFTJs. For example, three-component (or tricolor) superlattice (TCS) is one of the strategies to make artificial MFTJs. The theoretical prediction [249-251] and successful experimental realization [252] of asymmetric TCSs, e.g., BTO-CaTiO$_3$(CTO)-SrTiO$_3$(STO), set up good examples for demonstrating the power of the first-principles calculations to design this kind of MFTJ. The breaking of compositional inversion symmetry not only naturally brings asymmetric ferroelectricity into the system, but also provides additional freedom to tune the average lattice parameter. Similar spirit was presented in the work done in 2002 by Yamada et al. [253]. Yamada et al. [253] reported that the inversion symmetry could be broken by interface effects in artificial perovskite oxide tricolor superlattices, e.g., asymmetrically stacked (LaAlO$_3$/La$_{0.6}$Sr$_{0.4}$MnO$_3$/STO)$n$. Recently, Gao et al.'s [254] theoretical work thought that robust intrinsic asymmetric ferroelectricity could exist in TCSs of BTO-CTO-STO at the nanoscale. They also proved that these TCSs could still maintain stable structures at the interfaces contacting with ferromagnetic electrode materials such as SRO and the asymmetric ferroelectricity would not disappear in these MFTJs. As shown by asymmetric potential energy profiles of the soft mode distortions in Fig. 11, the three well-defined energy differences between "up" and "down" states clearly show the stack sequences and interface effect contribute to maintain the ferroelectricity in tricolor nanoscale superlattices. Such nanoscale intrinsic asymmetric ferroelectricity could be used in ferroelectric related tunnel junctions.

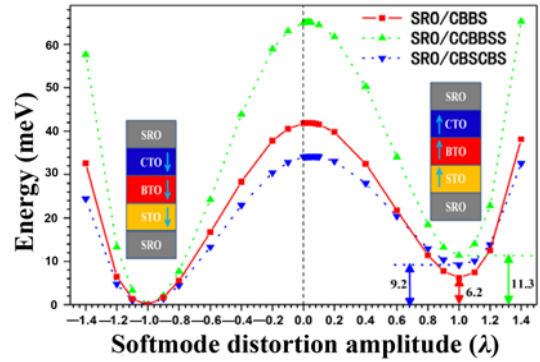

**Fig. 11** (Color online) Calculated asymmetric potential energy profiles of the soft mode distortions (characterized by $\lambda$, where $\lambda$= +1 and –1 correspond to the "up" and "down" states, respectively) in the SrRuO$_3$ (SRO) coververd tricolor lattices: SRO/CTO-(BTO)$_2$-STO (SRO/CBBS), SRO/(CTO)$_2$-(BTO)$_2$-(STO)$_2$ (SRO/CCBBSS,) and SRO/(CTO-BTO-STO)$_2$ (SRO/CBSCBS). The energy differences between "up" and "down"states are 6.2, 11.3 and 9.2 meV for CBBS, CCBBSS and CBSCBS, respectively. Reproduced with permission from [254]. Copyright 2013, Institute of Physics

### 3.4 Other types of electric control of magnetism

Charge and spin are the two basic properties of electrons. Compared with the conventional charge-based electronics, which has been extensively researched, the concept of spintronics (spin transport electronics or spin-based electronics) has been pointed out until the end of the 20th century [255]. As an attracting area of nanoscale electronics, the research of spintronics has experienced vigorous development in recent years.

The choice of functional materials for spintronics is nearly unlimited, such as semiconductors, metals, semiconductor/metal interface, transition metal oxides, and carbon nanostructure, etc. Due to the advantage of spintronics devices, i.e., quick spin response and less thermal energy loss, they would be major candidates for next-generation electronic devices with ultra-high speed and ultra-low power consumption [12].

The core problem of spintronics is how to switch the magnetization of a nanoscale memory element? In order to deal with it, a series of phenomena and devices related to spintronics have been discovered. Spin-polarized current control, such as racetrack memory based on the spin transfer torque (STT) effects, the most conventional magnetic field control, such as topological insulators based on the quantum spin Hall effect (QSHE), and photonic field control, such as lateral spin valve based on all-optical control in semiconductors, are the three main branches of it [256].

Very recently, as an entirely new branch, all-electric spintronics, which manipulate magnetic moments or electron spin states via electric fields, has received a great deal of attention. The representative examples include electric field control of magnetic anisotropy [214, 219, 223], and its control of ferromagnetism in multiferroics [157, 257]. Compared with magnetic field control, all-electric spintronics is a low-power magnetization switching technology, which is the main developing trend of spintronics and has the potential application in low-power logic devices and nonvolatile, solid-state memory. Above we have discussed direct electric field or ferroelectric control of MAE. Based on that it is theoretically possible to switch the magnetization solely by electric field [258, 259]. Following we will discuss other types of electric control of magnetism.

### 3.4.1 Rashba spin-orbit coupling

The SOC [260] describes an interaction between a particle′s spin and its motion, which has attracted considerable research attention in the field of spintronics. For the electron, SOC acts as an effective magnetic field, which makes spin of the electron respond to its orbital environment, thus all-electric manipulation of the spin degree of freedom is possible. Among various kinds of SOCs, Rashba-type SOC [261] has attracted the most research focus because of its tunability through external electric field, which has been experimentally implemented in semiconductor heterostructure [262]. Rashba SOC arises from the structure inversion asymmetry, and thus usually appears at the interface/surface. Its Hamiltonian can be written as:

$$H_R = \frac{\hbar}{4m^2c^2}(\nabla V \times p)\cdot \vec{\sigma}, \tag{4}$$

where $p$ is the momentum, $\sigma$ is the Pauli operator, and $\nabla V$ is the potential gradient at the surface/interface. Electron affected by this effect will have a spin direction perpendicular to momentum (Fig. S14).

### 3.4.2 Rashba spin-orbit coupling in semiconductors

The use of Rashba SOC in semiconductors for spintronics can be traced back at least to the theoretical proposal of a spin field-effect-transistor (FET) by Datta and Das [263] in 1990. Fig. S15 shows the schematic of Datta-Das spin FET, in which InAlAs/InGaAs heterostructure provides a two-dimensional channel for electron transport between two ferromagnetic electrodes, and Rashba SOC acts as a *k*-dependent effective magnetic field. In this device, the gate voltage is applied to tune the strength of Rashba SOC in the channel, which controls spin's precession and thus determines the transport current. This is a very novel spintronics device, however, due to various difficulties, for example, the low spin injection efficiency from semiconductor to ferromagnetic metal, the serious spin decoherence effect induced by strong SOC, etc., it still remains a theoretical model, and has yet to be confirmed experimentally. Although this Rashba SOC controlled spin FET has not been experimentally confirmed so far, it is regarded as one of the most advanced applications of spintronics in the future.

When a conductor/semiconductor is subject to orthogonal electric and magnetic fields, opposite charges will accumulate at the edges of a conductor/semiconductor due to the Lorentz force, which is known as the conventional Hall effect. Analogous to the conventional Hall effect, the spin Hall effect (SHE) refers to spin accumulation as a result of a charge current in the presence of SOC [264]. Such process is represented schematically in Fig. 12a. Two different types of SHE, i.e., intrinsic and extrinsic, have been recently proposed and observed in experiments. The intrinsic effect is caused by SOC in the band structure of the semiconductor and is independent on disorder [265], whereas the extrinsic SHE relies on scattering by impurities, typically includes the skew scattering and side-jump [266]. Using the Kerr effect, Kato et al. [267] observed the SHE in *n*-doped GaAs (Fig. 12b). Their observation is generally regarded as extrinsic SHE i.e., conduction electrons are scattered by impurities or defects in the crystal, where SOC causes the spin asymmetric effect. Another experimental report of the SHE was done by Wunderlich et al. [268]. They used a *p-n* junction light emitting diode (LED) device that coupled two-dimensional hole and electron doped systems with Rashba SOC (Fig. 12c). The detection of spin-polarized light was performed by observing the circularly polarized light emitted when the hole and electron combined.

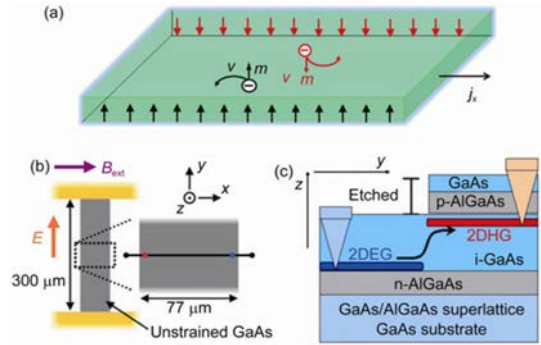

**Fig. 12** (Color online) **a** Spin accumulation is induced at the edges of the sample due to spin-orbit interaction when a pure charge current $j_x$ is applied [264]. **b** Schematic of the unstrained GaAs sample and the experimental geometry [267] **c** The schematic cross section of the coplanar p-n junction LED device. At forward bias of order of the GaAs band gap, electrons move from the 2DEG to the a two-dimensional hole gas (2DHG) where they recombine. The highest intensity of the emitted light is in the p region near the junction step edge[268]. Picture reproduced with permission from [269]. Copyright 2013, Springer

### 3.4.3 Rashba spin splitting at non-magnetic or magnetic metal surface

Metal surfaces form a new family for investigation of Rashba SOC. Research interest in metal surface begins with the Au(111) surface, in which a considerable Rashba spin splitting, about 110 meV at the Fermi level, was obtained in the *sp*-derived surface states through angle-resolved photoelectron spectroscopy [270]. Recent studies have extended to surface alloys, such as Ag/Au(111) [271], Ag/Pt(111) [272], Bi/Ag(111) [273], etc., aiming to tune the magnitude of the Rashba splitting strength by surface modification. The electric field, which is the most popular approach to tune Rashba SOC strength in semiconductor heterostructures, however, is seldom exploited in the metal surfaces. Bihlmayer et al. [274] were the first group to try the electric field control of Rashba splitting at a metal surface, yet they just provided very simple discussion. Very recently, Park et al. [275] calculated the Rashba splitting in a single Bi layer under an external electric field. A single metal layer itself, however, shows no inversion asymmetry, and thus an extremely strong electric field is required to produce considerable Rashba splitting, which may be of little practical significance. In 2013, Gong et al. [276] investigated the electric field control of Rashba spin splitting in Au(111) surface. Their calculations show that the electric field can modify the surface electrostatic potential and its gradient, resulting the shift of the Rashba splitting bands and change of the Rashba SOC strength. Fig. S16 shows the linear relationship between the Rashba splitting energy and the applied electric field.

Controlling the magnetic anisotropy of thin ferromagnetic films using a static electric field is of great interest and has been previously investigated, For example, Duan et al.'s [66] investigated the Fe, Co, Ni thin films imposed to the external electric field, and found the linear relationship between the surface MAE and the electric field. They omitted the contribution of the surface Rashba SOC, and attributed such electric field control of MAE to spin-dependent screening effect. For magnets, especially those with high-Z atoms, for example, in surface state of ferromagnetic Yb [277], Rashba SOC and exchange field coexist, and they interact with each other. Barnes et al. [278]developed a simple analytic theory for the magnetic surface, and demonstrated the influence of Rashba SOC on MAE. Fig. S17 schematically shows the band structure tuned by Rashba SOC or/and ferromagnetic exchange field.

3.4.4 Manipulation of magnetic properties through charge injection

Using relativistic density-functional calculations, Gong et al. [279, 280] calculated graphene/3$d$ complex systems. They investigated MAE of freestanding Fe monolayer and Fe/graphene complex system, and found MAE of Fe atom wass drastically changed, from meV/atom scale in freestanding Fe monolayer to l eV/atom in Fe/graphene system [280]. The most important finding is that, through charge injection, the suppressed MAE of Fe atoms in Fe/graphene system can be restored back, which provides an effective approach to control MAE.

Subsequently, Stepanyuk et al. [281] also used the method of charge injection to tune MAE of Fe-Pt multilayers, and found that MAE and the direction of magnetization in metallic magnetic multilayers can be tailored by surface charging. They demonstrated that surface charging has a deep effect on the magnitude of the MAE, which is composition dependent. Fig. S18a shows MAE versus the charge injection for systems featuring a single Fe layer capped with Pt. Fig. S18b, c show the behavior of the multilayers capped with iron bilayers, for which surface charging not only affects the value of the anisotropy but an easy-axis switching is also revealed.

**4 Summary and prospects**

Due to the magnetoelectric effect arising from the coupling between ferroelectricity and (anti)ferromagnetism, the multiferroics are significantly different from general ferroelectrics and ferromagnets. Hence multiferroic materials are unique systems in many ways and are expected to be incorporated into novel electronic devices and spintronics devices. However, multiferroics equipped with prominent performance at the room temperature are still limited.

In order to accelerate the discovery of novel materials, constructing an information library of materials is necessary. Materials Genome Initiative emerged at the right moment in 2011 which was launched by the US President Obama. Deriving from the concept of biological genome, the so-called "Materials Genome" is a set of comprehensive databases based on crystallographic unit cell of crystalline phases. As elemental composition, structural symmetry, and intrinsic properties can be conveniently reflected through unit cell, the Material Genome would be a powerful tool to predict a material's structure, physical/chemical properties, and even its response to processing and usage conditions.

Considering the potential application of first-principles calculations in optimization for lattice configuration, elastic, electronic, dielectric and thermodynamic properties, first-principles studies have been carried out in the establishment of the Materials Genome in the past few years [282]. With rapid enhancement in computer capabilities, they will play more and more important role in the prospective research of the Material Genome, including identifying the genome of multiferroics.

The implement of Material Genome demands numerous computations over kinds of materials. Traditional computing methods are powerless for this arduous task, but high throughput computing—an emerging technique seems most possibly to curb the problem. In the frame of high throughput computing based on first-principles theory or high throughput first-principles calculation, loosely coupled parallel computations for lots of materials are distributed on large amounts of computational resources over periods of time. Very recently, computational software integration [283], large-scale data storage & processing technology [284] and pseudopotentials library [285] for high throughput first-principles calculation have started to explode into a prominence role in screening cathode materials for Lithium-ion batteries [286] and high-performance piezoelectrics [287]. Though high throughput first-principles study of multiferroics has not set sail due to the technical barriers of themselves and the complexity of multiferroics, we are confident that a high-throughput highway for computational multiferroics design will be constructed and help to find satisfying multiferroics in coming years.

It is noticeable that a completely theoretical understanding of the behavior of multiferroics requires computational tools that span the atomic-scale detail of first-principles methods and the more coarse-grained description provided by continuum equations. Due to its superiority in dealing with grain boundary problems [288, 289], current effort for multiferroics research focuses on combining first-principles study, molecular dynamics, Monte Carlo simulation, spin models, and micromagnetic simulation into large unit multiscale framework. The realization of unit multiscale framework is a grand challenge for theoretical physicists, fortunately several recent cutting-edge work has made pioneering progress on acquiring the information of ferromagnetic or ferroelectric domain wall information [290, 291] and coherently controlling magnetization switching by a series of electric field pulse signals [292], which may motivate new thoughts to spring up in multiscale study of multiferroics.

Finally, it is difficult to achieve any information of multiferroics and to design multiferroic devices precisely without the support of materials growth technology, materials characterization & detection technique, and layout design in experiment. In consequence, first-principles study combining state-of-art experimental equipment, e.g., scanning tunneling microscope, atomic force microscopy, micro-Raman spectroscopy and transmission electron microscopy, now is becoming one of the most popular and hottest way to investigate properties of multiferroics [103, 104].

In conclusion, many efforts of experimental and theoretical investigations have been made to study multiferroics. There still exist, however, profound unknown physics blinding us to make use of multiferroic materials in real multi-level memories, sensors or transducers, *etc*. Recently, lots of motivational research combining first-principles study and experiments [104, 293, 294] shed lights on our way to realize materials design and devices fabrication. First-principles approach is now considerably matured and does not show any sign of fading in the research of multiferroics. In fact, due to the vast currently unexplored opportunity to tuning the properties of the systems by strain engineering or structural modifications, first-principles study of multiferroics will continue to stand out in the arsenal of condensed matter physics.

# Acknowledgment

This work was supported by the National Basic Research Program of China (2014CB921104, 2013CB922301), the National Natural Science Foundation of China (61125403), Program of Shanghai Subject Chief Scientist, and Fundamental Research Funds for the central universities (ECNU). Hing-Chen Ding acknowledges the support from ECNU-PY2012001.

# References

1. Curie P (1894) Sur la symétrie dans les phénomènes physiques, symétrie d'un champ électrique et d'un champ magnétique. J Phys Theor Appl 3: 393-415
2. Schmid H (1994) Multi-ferroic magnetoelectrics. Ferroelectrics 162: 317-338
3. Fiebig M (2005) Revival of the magnetoelectric effect. J Phys D: Appl Phys 38: R123-R152
4. Ramesh R, Spaldin NA (2007) Multiferroics: Progress and prospects in thin films. Nat Mater 6: 21-29
5. Cheong S-W, Mostovoy M (2007) Multiferroics: A magnetic twist for ferroelectricity. Nat Mater 6: 13-20
6. Nan C-W, Bichurin MI, Dong S, et al (2008) Multiferroic magnetoelectric composites: Historical perspective, status, and future directions. J Appl Phys 103: 031101
7. Wang KF, Liu JM, Ren ZF (2009) Multiferroicity: The coupling between magnetic and polarization orders. Adv Phys 58: 321-448
8. Spaldin NA, Cheong S-W, Ramesh R (2010) Multiferroics: Past, present, and future. Phys Today 63: 38-43
9. Ma J, Hu J, Li Z, et al (2011) Recent progress in multiferroic magnetoelectric composites: From bulk to thin films. Adv Mater 23: 1062-1087
10. Scott JF (2012) Applications of magnetoelectrics. J Mater Chem 22: 4567-4574
11. Duan C-G (2012) Interface/surface magnetoelectric effects: New routes to the electric field control of magnetism. Front Phys 7: 375-379
12. Gong S, Ding H, Zhu W, et al (2013) A new pathway towards all-electric spintronics: Electric-field control of spin states through surface/interface effects. Sci China Phys Mech Astron 56: 232-244
13. Hill NA (2002) Density functional studies of multiferroic magnetoelectrics. Annu Rev Mater Res 32: 1-37
14. Ederer C, Spaldin NA (2005) Recent progress in first-principles studies of magnetoelectric multiferroics. Curr Opin Solid State Mater Sci 9: 128-139
15. Silvia P, Claude E (2009) First principles studies of multiferroic materials. J Phys: Condens Matter 21: 303201
16. Milman V, Winkler B, White JA, et al (2000) Electronic structure, properties, and phase stability of inorganic crystals: A pseudopotential plane-wave study. Int J Quantum Chem 77: 895-910
17. Wentzcovitch RM, Martins JL, Allen PB (1992) Energy versus free-energy conservation in first-principles molecular dynamics. Phys Rev B 45: 11372-11374
18. Mujica A, Rubio A, Muñoz A, et al (2003) High-pressure phases of group-IV, III–V, and II–VI compounds. Rev Mod Phys 75: 863-912
19. Bousquet E, Spaldin NA, Ghosez P (2010) Strain-induced ferroelectricity in simple rocksalt binary oxides. Phys Rev Lett 104: 037601
20. Hirose K, Tsukada M (1995) First-principles calculation of the electronic structure for a bielectrode junction system under strong field and current. Phys Rev B 51: 5278-5290
21. Kobayashi KI, Kimura T, Tomioka Y, et al (1999) Intergrain tunneling magnetoresistance in polycrystals of the ordered double perovskite $Sr_2FeReO_6$. Phys Rev B 59: 11159-11162
22. Singh DJ (2010) Doping-dependent thermopower of PbTe from boltzmann transport calculations. Phys Rev B 81: 195217
23. Carlsson JM, Hellsing B (2000) First-principles investigation of the quantum-well system Na on Cu(111). Phys Rev B 61: 13973-13982
24. Read AJ, Needs RJ, Nash KJ, et al (1992) First-principles calculations of the electronic properties of silicon quantum wires. Phys Rev Lett 69: 1232-1235
25. Li J, Wang L-W (2004) First principle study of core/shell structure quantum dots. Appl Phys Lett 84: 3648-3650
26. Zhang H, Liu C-X, Qi X-L, et al (2009) Topological insulators in $Bi_2Se_3$, $Bi_2Te_3$ and $Sb_2Te_3$ with a single dirac cone on the surface. Nat Phys 5: 438-442
27. Butler WH, Zhang XG, Schulthess TC, et al (2001) Spin-dependent tunneling conductance of Fe|MgO|Fe sandwiches. Phys Rev B 63: 054416
28. Liu AY, Cohen ML (1989) Prediction of new low compressibility solids. Science 245: 841-842
29. Van de Walle CG, Neugebauer J (2004) First-principles calculations for defects and impurities: Applications to III-nitrides. J Appl Phys 95: 3851-3879
30. Demkov AA, Ortega J, Sankey OF, et al (1995) Electronic structure approach for complex silicas. Phys Rev B 52: 1618-1630
31. Van der Ven A, Aydinol MK, Ceder G, et al (1998) First-principles investigation of phase stability in $Li_xCoO_2$. Phys Rev B 58: 2975-2987
32. Tsuchiya T, Tsuchiya J, Umemoto K, et al (2004) Phase transition in MgSiO3 perovskite in the earth's lower mantle. Earth Planet Sci Lett 224: 241-248
33. Hatt AJ, Spaldin NA, Ederer C (2010) Strain-induced isosymmetric phase transition in $BiFeO_3$. Phys Rev B 81: 054109
34. Wang H, Wang B, Li QK, et al (2007) First-principles study of the cubic perovskites $BiMO_3$ (M=Al, Ga, In, and Sc). Phys Rev B 75: 245209
35. Ding H-C, Duan C-G (2012) Electric-field control of magnetic ordering in the tetragonal-like $BiFeO_3$. Europhys Lett 97: 57007
36. King-Smith RD, Vanderbilt D (1994) First-principles investigation of ferroelectricity in perovskite compounds. Phys Rev B 49: 5828-5844
37. Van Aken BB, Palstra TTM, Filippetti A, et al (2004) The origin of ferroelectricity in magnetoelectric $YMnO_3$. Nat Mater 3: 164-170
38. Neaton JB, Ederer C, Waghmare UV, et al (2005) First-principles study of spontaneous polarization in multiferroic $BiFeO_3$. Phys Rev B 71: 014113
39. Xu Y-N, Ching WY (1993) Electronic, optical, and structural properties of some wurtzite crystals. Phys Rev B 48: 4335-4351
40. Suzuki M, Uenoyama T, Yanase A (1995) First-principles calculations of effective-mass parameters of AlN and GaN. Phys Rev B 52: 8132-8139
41. Delaney KT, Mostovoy M, Spaldin NA (2009) Superexchange-driven magnetoelectricity in magnetic vortices. Phys Rev Lett 102: 157203

42. Eyert V (2002) The metal-insulator transitions of $VO_2$: A band theoretical approach. Ann Phys-Berlin 11: 650-704
43. Pavarini E, Biermann S, Poteryaev A, et al (2004) Mott transition and suppression of orbital fluctuations in orthorhombic $3d^1$ perovskites. Phys Rev Lett 92: 176403
44. Copple A, Ralston N, Peng X (2012) Engineering direct-indirect band gap transition in wurtzite gaas nanowires through size and uniaxial strain. Appl Phys Lett 100: 193108
45. Asahi R, Taga Y, Mannstadt W, et al (2000) Electronic and optical properties of anatase $TiO_2$. Phys Rev B 61: 7459-7465
46. Rollmann G, Rohrbach A, Entel P, et al (2004) First-principles calculation of the structure and magnetic phases of hematite. Phys Rev B 69: 165107
47. Nielsen OH, Martin RM (1983) First-principles calculation of stress. Phys Rev Lett 50: 697-700
48. Kim K, Lambrecht WRL, Segall B (1996) Elastic constants and related properties of tetrahedrally bonded BN, AlN, GaN, and InN. Phys Rev B 53: 16310-16326
49. Shang SL, Saengdeejing A, Mei ZG, et al (2010) First-principles calculations of pure elements: Equations of state and elastic stiffness constants. Comput Mater Sci 48: 813-826
50. Roundy D, Krenn CR, Cohen ML, et al (1999) Ideal shear strengths of fcc aluminum and copper. Phys Rev Lett 82: 2713-2716
51. Ogata S, Li J, Yip S (2002) Ideal pure shear strength of aluminum and copper. Science 298: 807-811
52. Telling RH, Pickard CJ, Payne MC, et al (2000) Theoretical strength and cleavage of diamond. Phys Rev Lett 84: 5160-5163
53. Roundy D, Cohen ML (2001) Ideal strength of diamond, Si, and Ge. Phys Rev B 64: 212103
54. Ogata S, Shibutani Y (2003) Ideal tensile strength and band gap of single-walled carbon nanotubes. Phys Rev B 68: 165409
55. Liu F, Ming P, Li J (2007) Ab initio calculation of ideal strength and phonon instability of graphene under tension. Phys Rev B 76: 064120
56. Eriksson O, Johansson B, Albers RC, et al (1990) Orbital magnetism in Fe, Co, and Ni. Phys Rev B 42: 2707-2710
57. Yazyev OV, Helm L (2007) Defect-induced magnetism in graphene. Phys Rev B 75: 125408
58. Jiang D-e, Sumpter BG, Dai S (2007) First principles study of magnetism in nanographenes. J Chem Phys 127: 124703
59. Yang K, Wu R, Shen L, et al (2010) Origin of $d^0$ magnetism in II-VI and III-V semiconductors by substitutional doping at anion site. Phys Rev B 81: 125211
60. Yang JY, Han YL, He L, et al (2012) D carrier induced intrinsic room temperature ferromagnetism in $Nb:TiO_2$ film. Appl Phys Lett 100: 202409
61. Tischer M, Hjortstam O, Arvanitis D, et al (1995) Enhancement of orbital magnetism at surfaces: Co on Cu(100). Phys Rev Lett 75: 1602-1605
62. Daalderop GHO, Kelly PJ, Schuurmans MFH (1990) First-principles calculation of the magnetocrystalline anisotropy energy of iron, cobalt, and nickel. Phys Rev B 41: 11919-11937
63. Wang D-s, Wu R, Freeman AJ (1993) State-tracking first-principles determination of magnetocrystalline anisotropy. Phys Rev Lett 70: 869-872
64. Wang D-s, Wu R, Freeman AJ (1993) First-principles theory of surface magnetocrystalline anisotropy and the diatomic-pair model. Phys Rev B 47: 14932-14947
65. Burkert T, Nordström L, Eriksson O, et al (2004) Giant magnetic anisotropy in tetragonal FeCo alloys. Phys Rev Lett 93: 027203
66. Duan C-G, Velev JP, Sabirianov RF, et al (2008) Surface magnetoelectric effect in ferromagnetic metal films. Phys Rev Lett 101: 137201
67. Nakamura K, Shimabukuro R, Fujiwara Y, et al (2009) Giant modification of the magnetocrystalline anisotropy in transition-metal monolayers by an external electric field. Phys Rev Lett 102: 187201
68. Baettig P, Ederer C, Spaldin NA (2005) First principles study of the multiferroics $BiFeO_3$, $Bi_2FeCrO_6$, and $BiCrO_3$: Structure, polarization, and magnetic ordering temperature. Phys Rev B 72: 214105
69. Han MJ, Ozaki T, Yu JJ (2007) Magnetic ordering and exchange interactions in multiferroic $GaFeO_3$. Phys Rev B 75: 060404
70. Okada S, Oshiyama A (2001) Magnetic ordering in hexagonally bonded sheets with first-row elements. Phys Rev Lett 87: 146803
71. Lee H, Son Y-W, Park N, et al (2005) Magnetic ordering at the edges of graphitic fragments: Magnetic tail interactions between the edge-localized states. Phys Rev B 72: 174431
72. Dal D, Whangbo MH, Koo HJ, et al (2005) Analysis of the spin exchange interactions and the ordered magnetic structures of lithium transition metal phosphates $LiMPO_4$ (M = Mn, Fe, Co, Ni) with the olivine structure. Inorg Chem 44: 2407-2413
73. Velev JP, Duan C-G, Belashchenko KD, et al (2007) Effect of ferroelectricity on electron transport in $Pt/BaTiO_3/Pt$ tunnel junctions. Phys Rev Lett 98: 137201
74. Wang J, Neaton JB, Zheng H, et al (2003) Epitaxial $BiFeO_3$ multiferroic thin film heterostructures. Science 299: 1719-1722
75. Fu H, Cohen RE (2000) Polarization rotation mechanism for ultrahigh electromechanical response in single-crystal piezoelectrics. Nature 403: 281-283
76. Ederer C, Spaldin NA (2005) Influence of strain and oxygen vacancies on the magnetoelectric properties of multiferroic bismuth ferrite. Phys Rev B 71: 224103
77. Baettig P, Spaldin NA (2005) Ab initio prediction of a multiferroic with large polarization and magnetization. Appl Phys Lett 86: 012505
78. Lee JH, Rabe KM (2010) Epitaxial-strain-induced multiferroicity in $SrMnO_3$ from first principles. Phys Rev Lett 104: 207204
79. Ederer C, Spaldin NA (2005) Effect of epitaxial strain on the spontaneous polarization of thin film ferroelectrics. Phys Rev Lett 95: 257601
80. Waghmare UV, Rabe KM (1997) Ab initio statistical mechanics of the ferroelectric phase transition in $PbTiO_3$. Phys Rev B 55: 6161-6173
81. Noheda B, Cox DE, Shirane G, et al (2001) Polarization rotation via a monoclinic phase in the piezoelectric 92% $PbZn_{1/3}Nb_{2/3}O_3$-8% $PbTiO_3$. Phys Rev Lett 86: 3891-3894
82. Wu Z, Krakauer H (2003) First-principles calculations of piezoelectricity and polarization rotation in $Pb(Zr_{0.5}Ti_{0.5})O_3$. Phys Rev B 68: 014112
83. Meyer B, Vanderbilt D (2002) *Ab initio* study of ferroelectric domain walls in $PbTiO_3$. Phys Rev B 65: 104111
84. Seidel J, Martin LW, He Q, et al (2009) Conduction at domain walls in oxide multiferroics. Nat Mater 8: 229-234
85. Kohn W, Sham LJ (1965) Self-consistent equations including exchange and correlation effects. Phys Rev 140: A1133-A1138
86. Aryasetiawan F, Gunnarsson O (1998) The GW method. Rep Prog Phys 61: 237
87. Rohlfing M, Louie SG (2000) Electron-hole excitations and optical spectra from first principles. Phys Rev B 62: 4927-4944
88. Benedict LX, Shirley EL, Bohn RB (1998) Optical absorption of insulators and the electron-hole interaction: An *ab initio* calculation. Phys Rev Lett 80: 4514-4517
89. Li J, Duan C-G, Gu Z-q, et al (1998) First-principles calculations of the electronic structure and optical properties of $LiB_3O_5$, $CsB_3O_5$, and $BaB_2O_4$ crystals. Phys Rev B 57: 6925-6932
90. Saha S, Sinha TP, Mookerjee A (2000) Electronic structure, chemical bonding, and optical properties of paraelectric $BaTiO_3$. Phys Rev B 62: 8828-8834
91. Ambrosch-Draxl C, Majewski JA, Vogl P, et al (1995) First-principles studies of the structural and optical properties of crystalline poly(*para*-phenylene). Phys Rev B 51: 9668-9676
92. Skorodumova NV, Ahuja R, Simak SI, et al (2001) Electronic, bonding, and optical properties of $CeO_2$ and $Ce_2O_3$ from first principles. Phys Rev B 64: 115108 93  Yang L, Deslippe J, Park C-H, et al (2009) Excitonic effects on the optical response of graphene and bilayer graphene. Phys Rev Lett 103: 186802
94  Ghosh DB, De M, De SK (2004) Electronic structure and magneto-optical properties of magnetic semiconductors: Europium monochalcogenides. Phys Rev B 70: 115211
95  Antonov VN, Harmon BN, Antropov VP, et al (2001) Electronic structure and magneto-optical kerr effect of $Fe_3O_4$ and $Mg^{2+}$- or $Al^{3+}$-substituted $Fe_3O_4$. Phys Rev B 64: 134410
96  Kuneš J, Novák P, Diviš M, et al (2001) Magnetic, magneto-optical, and structural properties of URhAl from first-principles calculations. Phys Rev B 63: 205111
97  Duan C-G, Li J, Gu Z-q, et al (1999) Interpretation of the nonlinear optical susceptibility of borate crystals from first principles. Phys Rev B 59: 369-372
98  Duan C-G, Li J, Gu Z-q, et al (1999) First-principles calculation of the second-harmonic-generation coefficients of berate crystals. Phys Rev B 60: 9435-9443
99  Mendoza BS, Palummo M, Onida G, et al (2001) *Ab initio* calculation of second-harmonic-generation at the Si(100) surface. Phys Rev B 63: 205406
100  Rashkeev SN, Lambrecht WRL (2001) Second-harmonic generation of I-III-$VI_2$ chalcopyrite semiconductors: Effects of chemical substitutions. Phys Rev B 63: 165212
101  Tong W-Y, Ding H-C, Gao Y-C, et al (2014) Spin-dependent optical response of multiferroic EuO: First-principles DFT calculations. Phys Rev B 89: 064404
102  Lee JS, Xie YW, Sato HK, et al (2013) Titanium $d_{xy}$ ferromagnetism at the $LaAlO_3/SrTiO_3$ interface. Nat Mater 12: 703-706
103  Zhang J, Ke X, Gou G, et al (2013) A nanoscale shape memory oxide. Nat Commun 4: 2768
104  Huang R, Ding H-C, Liang W-I, et al (2014) Atomic-scale visualization of polarization pinning and relaxation at coherent $BiFeO_3/LaAlO_3$ interfaces. Adv Funct Mater 24: 793-799
105  Zhu W, Ding H-C, Gong S-J, et al (2013) First-principles studies of the magnetic anisotropy of the Cu/FePt/MgO system. J Phys: Condens Matter 25: 396001
106  Kroemer H (2001) Nobel lecture: Quasielectric fields and band offsets: Teaching electrons new tricks. Rev Mod Phys 73: 783-793
107  Di Ventra M, Pantelides ST, Lang ND (2000) First-principles calculation of transport properties of a molecular device. Phys Rev Lett 84: 979-982
108  Junquera J, Ghosez P (2003) Critical thickness for ferroelectricity in perovskite ultrathin films. Nature 422: 506-509
109  Duan C-G, Sabirianov RF, Mei W-N, et al (2006) Interface effect on ferroelectricity at the nanoscale. Nano Lett 6: 483-487
110  Duan C-G, Jaswal SS, Tsymbal EY (2006) Predicted magnetoelectric effect in $Fe/BaTiO_3$ multilayers: Ferroelectric control of magnetism. Phys Rev Lett 97: 047201
111  Dalverny AL, Filhol JS, Doublet ML (2011) Interface electrochemistry in conversion materials for Li-ion batteries. J Mater Chem 21: 10134-10142
112  Liu H, Neal AT, Zhu Z, et al (2014) Phosphorene: An unexplored 2D semiconductor with a high hole mobility. ACS Nano 8: 4033-4041
113  Iijima S (1991) Helical microtubules of graphitic carbon. Nature 354: 3
114  Hashimoto A, Suenaga K, Gloter A, et al (2004) Direct evidence for atomic defects in graphene layers. Nature 430: 870-873
115  Vogt P, De Padova P, Quaresima C, et al (2012) Silicene: Compelling experimental evidence for graphenelike two-dimensional silicon. Phys Rev Lett 108: 155501
116  Hong SS, Cha JJ, Kong D, et al (2012) Ultra-low carrier concentration and surface-dominant transport in antimony-doped $Bi_2Se_3$ topological insulator nanoribbons. Nat Commun 3: 757
117  Charlier J-C, Blase X, Roche S (2007) Electronic and transport properties of nanotubes. Rev Mod Phys 79: 677-732
118  Guo W (2013) Do silicene nanoribbons have high carrier mobilities? Europhys Lett 101: 27005
119  Gao J, Zhao J (2012) Initial geometries, interaction mechanism and high stability of silicene on Ag(111) surface. Sci Rep 2: 861
120  Nemec L, Blum V, Rinke P, et al (2013) Thermodynamic equilibrium conditions of graphene films on SiC. Phys Rev Lett 111: 065502
121  Kudin KN, Ozbas B, Schniepp HC, et al (2007) Raman spectra of graphite oxide and functionalized graphene sheets. Nano Lett 8: 36-41
122  Li Z, Shao S, Li N, et al (2013) Single crystalline nanostructures of topological crystalline insulator SnTe with distinct facets and morphologies. Nano Lett 13: 5443-5448
123  Peng S, Cho K (2003) Ab initio study of doped carbon nanotube sensors. Nano Lett 3: 513-517
124  Tsai W-F, Huang C-Y, Chang T-R, et al (2013) Gated silicene as a tunable source of nearly 100% spin-polarized electrons. Nat Commun 4: 1500
125  Tong Z, Liang Y, Jiang X, et al (2014) An atomistic investigation on the mechanism of machining nanostructures when using single tip and multi-tip diamond tools. Appl Surf Sci 290: 458-465
126  Sharifi-Mood N, Koplik J, Maldarelli C (2013) Molecular dynamics simulation of the motion of colloidal nanoparticles in a solute concentration gradient and a comparison to the continuum limit. Phys Rev Lett 111: 184501
127  Zhang K, Stocks GM, Zhong J (2007) Melting and premelting of carbon nanotubes. Nanotechnology 18: 285703
128  Tan X, Wu J, Zhang K, et al (2013) Nanoindentation models and young's modulus of monolayer graphene: A molecular dynamics study. Appl Phys Lett 102: 071908
129  Zheng M, Pavan GM, Neeb M, et al (2012) Targeting the blind spot of polycationic nanocarrier-based sirna delivery. ACS Nano 6: 9447-9454
130  Tou WI, Chang S-S, Lee C-C, et al (2013) Drug design for neuropathic pain regulation from traditional chinese medicine. Sci Rep 3: 844
131  Car R, Parrinello M (1985) Unified approach for molecular dynamics and density-functional theory. Phys Rev Lett 55: 2471-2474
132  Hutter J (2012) Car–parrinello molecular dynamics. Wiley Interdisciplinary Reviews: Computational Molecular Science 2: 604-612
133  Martoňák R, Laio A, Parrinello M (2003) Predicting crystal structures: The Parrinello-Rahman method revisited. Phys Rev Lett 90: 075503
134  Jungwirth P, Tobias DJ (2002) Ions at the air/water interface. J Phys Chem B 106: 6361-6373
135  Kuo IFW, Mundy CJ, McGrath MJ, et al (2004) Liquid water from first principles: Investigation of different sampling approaches. J Phys Chem B 108: 12990-12998
136  Iannuzzi M, Laio A, Parrinello M (2003) Efficient exploration of reactive potential energy surfaces using Car-Parrinello molecular dynamics. Phys Rev Lett 90: 238302
137  Gavillet J, Loiseau A, Journet C, et al (2001) Root-growth mechanism for single-wall carbon nanotubes. Phys Rev Lett 87: 275504
138  Wdowik UD, Parlinski K (2007) Lattice dynamics of CoO from first principles. Phys Rev B 75: 104306
139  Ye L-H, Asahi R, Peng L-M, et al (2012) Model GW study of the late transition metal monoxides. J Chem Phys 137: 154110
140  Jakobi E, Kanungo S, Sarkar S, et al (2011) LDA+DMFT study of Ru-based perovskite $SrRuO_3$ and $CaRuO_3$. Phys Rev B 83: 041103
141  Cossu F, Schwingenschlögl U, Eyert V (2013) Metal-insulator transition at the $LaAlO_3/SrTiO_3$ interface revisited: A hybrid functional study. Phys Rev B 88: 045119
142  Souvi SMO, Sharkas K, Toulouse J (2014) Double-hybrid density-functional theory with meta-generalized-gradient approximations. J Chem Phys 140: 084107
143  Jiang H (2010) The GW method: Basic principles, latest developments and its applications for d- and f- electron systems. Acta Phys-Chim Sin 26:

144  Choly N, Kaxiras E (2002) Kinetic energy density functionals for non-periodic systems. Solid State Commun 121: 281-286
145  Spaldin NA (2012) A beginner's guide to the modern theory of polarization. J Solid State Chem 195: 2-10
146  King-Smith RD, Vanderbilt D (1993) Theory of polarization of crystalline solids. Phys Rev B 47: 1651-1654
147  Thonhauser T, Ceresoli D, Vanderbilt D, et al (2005) Orbital magnetization in periodic insulators. Phys Rev Lett 95: 137205
148  Anisimov VI, Zaanen J, Andersen OK (1991) Band theory and mott insulators: Hubbard U instead of stoner i. Phys Rev B 44: 943-954
149  Dudarev SL, Botton GA, Savrasov SY, et al (1998) Electron-energy-loss spectra and the structural stability of nickel oxide:An LSDA+U study. Phys Rev B 57: 1505-1509
150  Du Y, Ding H-C, Sheng L, et al (2014) Microscopic origin of stereochemically active lone pair formation from orbital selective external potential calculations. J Phys: Condens Matter 26: 025503
151  Hill NA (2000) Why are there so few magnetic ferroelectrics? J Phys Chem B 104: 6694-6709
152  Gajek M, Bibes M, Fusil S, et al (2007) Tunnel junctions with multiferroic barriers. Nat Mater 6: 296-302
153  Israel C, Mathur ND, Scott JF (2008) A one-cent room-temperature magnetoelectric sensor. Nat Mater 7: 93-94
154  Velev JP, Duan C-G, Burton JD, et al (2009) Magnetic tunnel junctions with ferroelectric barriers: Prediction of four resistance states from first principles. Nano Lett 9: 427-432
155  Scott JF (2007) Data storage - multiferroic memories. Nat Mater 6: 256-257
156  Bibes M, Barthelemy A (2008) Multiferroics: Towards a magnetoelectric memory. Nat Mater 7: 425-426
157  Chu Y-H, Martin LW, Holcomb MB, et al (2008) Electric-field control of local ferromagnetism using a magnetoelectric multiferroic. Nat Mater 7: 478-482
158  Zavaliche F, Zhao T, Zheng H, et al (2007) Electrically assisted magnetic recording in multiferroic nanostructures. Nano Lett 7: 1586-1590
159  Annette B-H (2012) The polarizability model for ferroelectricity in perovskite oxides. J Phys: Condens Matter 24: 273202
160  Cochran W (1959) Crystal stability and the theory of ferroelectricity. Phys Rev Lett 3: 412-414
161  Anderson PW (1960) Fizika dielektrikov. Moscow: Acad. Nauk, SSSR:
162  Barker AS, Tinkham M (1962) Far-infrared ferroelectric vibration mode in $SrTiO_3$. Phys Rev 125: 1527-1530
163  Dupé B, Prosandeev S, Geneste G, et al (2011) $BiFeO_3$ films under tensile epitaxial strain from first principles. Phys Rev Lett 106: 237601
164  Fan Z, Wang J, Sullivan MB, et al (2014) Structural instability of epitaxial (001) $BiFeO_3$ thin films under tensile strain. Sci Rep 4: 4631
165  Wei S-H, Krakauer H (1985) Local-density-functional calculation of the pressure-induced metallization of BaSe and BaTe. Phys Rev Lett 55: 1200-1203
166  Hedin L, Lundqvist BI (1971) Explicit local exchange-correlation potentials. J Phys C: Solid State Phys 4: 2064
167  Cohen RE, Krakauer H (1990) Lattice dynamics and origin of ferroelectricity in $BaTiO_3$: Linearized-augmented-plane-wave total-energy calculations. Phys Rev B 42: 6416-6423
168  Cohen RE (1992) Origin of ferroelectricity in perovskite oxides. Nature 358: 136-138138
169  Wan X, Ding H-C, Savrasov SY, et al (2012) Ferroelectricity induced by interatomic magnetic exchange interaction. arXiv:12023381
170  Seshadri R, Hill NA (2001) Visualizing the role of bi 6s "lone pairs" in the off-center distortion in ferromagnetic $BiMnO_3$. Chem Mater 13: 2892-2899
171  Fennie CJ, Rabe KM (2005) Ferroelectric transition in $YMnO_3$ from first principles. Phys Rev B 72: 100103
172  Yang CH, Seidel J, Kim SY, et al (2009) Electric modulation of conduction in multiferroic Ca-doped $BiFeO_3$ films. Nat Mater 8: 485-493
173  Yu P, Lee JS, Okamoto S, et al (2010) Interface ferromagnetism and orbital reconstruction in $BiFeO_3$-$La_{0.7}Sr_{0.3}MnO_3$ heterostructures. Phys Rev Lett 105: 027201
174  Zeches RJ, Rossell MD, Zhang JX, et al (2009) A strain-driven morphotropic phase boundary in $BiFeO_3$. Science 326: 977-980
175  Catalan G, Scott JF (2009) Physics and applications of bismuth ferrite. Adv Mater 21: 2463-2485
176  Kubel F, Schmid H (1990) Structure of a ferroelectric and ferroelastic monodomain crystal of the perovskite $BiFeO_3$. Acta Crystallogr Sect B: Struct Sci B46: 698-702
177  Ederer C, Spaldin NA (2005) Weak ferromagnetism and magnetoelectric coupling in bismuth ferrite. Phys Rev B 71: 060401
178  Zhang JX, He Q, Trassin M, et al (2011) Microscopic origin of the giant ferroelectric polarization in tetragonal-like $BiFeO_3$. Phys Rev Lett 107: 147602
179  Zhao T, Scholl A, Zavaliche F, et al (2006) Electrical control of antiferromagnetic domains in multiferroic $BiFeO_3$ films at room temperature. Nat Mater 5: 823-829
180  Catalan G, Seidel J, Ramesh R, et al (2012) Domain wall nanoelectronics. Rev Mod Phys 84: 119-156
181  Ren W, Yang Y, Diéguez O, et al (2013) Ferroelectric domains in multiferroic $BiFeO_3$ films under epitaxial strains. Phys Rev Lett 110: 187601
182  Fennie CJ, Rabe KM (2006) Magnetic and electric phase control in epitaxial $EuTiO_3$ from first principles. Phys Rev Lett 97: 267602
183  Lee JH, Fang L, Vlahos E, et al (2010) A strong ferroelectric ferromagnet created by means of spin-lattice coupling. Nature 466: 954-U972
184  Ingle NJC, Elfimov IS (2008) Influence of epitaxial strain on the ferromagnetic semiconductor EuO: First-principles calculations. Phys Rev B 77: 121202
185  Benedek NA, Fennie CJ (2011) Hybrid improper ferroelectricity: A mechanism for controllable polarization-magnetization coupling. Phys Rev Lett 106: 107204
186  Harris AB (2011) Symmetry analysis for the Rudlesden-Popper systems $Ca_3Mn_2O_7$ and $Ca_3Ti_2O_7$. Phys Rev B 84: 064116
187  Schlom DG, Chen L-Q, Eom C-B, et al (2007) Strain tuning of ferroelectric thin films. Annu Rev Mater Res 37: 589-626
188  Yang C-H, Kan D, Takeuchi I, et al (2012) Doping $BiFeO_3$: Approaches and enhanced functionality. Phys Chem Chem Phys 14: 15953-15962
189  Nechache R, Harnagea C, Carignan LP, et al (2007) Epitaxial $Bi_2FeCrO_6$ multiferroic thin films. Philos Mag Lett 87: 231-240
190  Kanamori J (1959) Superexchange interaction and symmetry properties of electron orbitals. J Phys Chem Solids 10: 87-98
191  Ueda K, Tabata H, Kawai T (1998) Ferromagnetism in $LaFeO_3$-$LaCrO_3$ superlattices. Science 280: 1064-1066
192  Palova L, Chandra P, Rabe KM (2010) Magnetostructural effect in the multiferroic $BiFeO_3$-$BiMnO_3$ checkerboard from first principles. Phys Rev Lett 104: 037202
193  Kimura T, Goto T, Shintani H, et al (2003) Magnetic control of ferroelectric polarization. Nature 426: 55-58
194  Goto T, Kimura T, Lawes G, et al (2004) Ferroelectricity and giant magnetocapacitance in perovskite rare-earth manganites. Phys Rev Lett 92: 257201
195  Prokhnenko O, Feyerherm R, Dudzik E, et al (2007) Enhanced ferroelectric polarization by induced Dy spin order in multiferroic $DyMnO_3$. Phys Rev Lett 98: 057206
196  White JS, Bator M, Hu Y, et al (2013) Strain-induced ferromagnetism in antiferromagnetic $LuMnO_3$ thin films. Phys Rev Lett 111: 037201
197  Hur N, Park S, Sharma PA, et al (2004) Electric polarization reversal and memory in a multiferroic material induced by magnetic fields. Nature 429:

392-395

198 Chapon LC, Blake GR, Gutmann MJ, et al (2004) Structural anomalies and multiferroic behavior in magnetically frustrated TbMn$_2$O$_5$. Phys Rev Lett 93: 177402

199 Wang C, Guo GC, He L (2007) Ferroelectricity driven by the noncentrosymmetric magnetic ordering in multiferroic TbMn$_2$O$_5$: A first- principles study. Phys Rev Lett 99: 177202

200 Sergienko IA, Dagotto E (2006) Role of the Dzyaloshinskii-Moriya interaction in multiferroic perovskites. Phys Rev B 73: 094434

201 Katsura H, Nagaosa N, Balatsky AV (2005) Spin current and magnetoelectric effect in noncollinear magnets. Phys Rev Lett 95: 057205

202 Xiang HJ, Kan EJ, Zhang Y, et al (2011) General theory for the ferroelectric polarization induced by spin-spiral order. Phys Rev Lett 107: 157202

203 Kurumaji T, Seki S, Ishiwata S, et al (2011) Magnetic-field induced competition of two multiferroic orders in a triangular-lattice helimagnet mni$_2$. Phys Rev Lett 106: 167206

204 Lu XZ, Whangbo MH, Dong S, et al (2012) Giant ferroelectric polarization of CaMn$_7$O$_{12}$ induced by a combined effect of Dzyaloshinskii-Moriya interaction and exchange striction. Phys Rev Lett 108: 187204

205 Seki S, Yamasaki Y, Soda M, et al (2008) Correlation between spin helicity and an electric polarization vector in quantum-spin chain magnet LiCu$_2$O$_2$. Phys Rev Lett 100: 127201

206 Lukashev PV, Paudel TR, López-Encarnación JM, et al (2012) Ferroelectric control of magnetocrystalline anisotropy at cobalt/poly(vinylidene fluoride) interfaces. Acs Nano 6: 9745-9750

207 Meyerheim HL, Klimenta F, Ernst A, et al (2011) Structural secrets of multiferroic interfaces. Phys Rev Lett 106: 087203

208 Niranjan MK, Velev JP, Duan CG, et al (2008) Magnetoelectric effect at the Fe$_3$O$_4$/BaTiO$_3$ (001) interface: A first-principles study. Phys Rev B 78: 104405

209 Park MS, Song J-H, Freeman AJ (2009) Charge imbalance and magnetic properties at the Fe$_3$O$_4$/BaTiO$_3$ interface. Phys Rev B 79: 024420

210 Zheng H, Wang J, Lofland SE, et al (2004) Multiferroic BaTiO$_3$-CoFe$_2$O$_4$ nanostructures. Science 303: 661-663

211 Fechner M, Maznichenko IV, Ostanin S, et al (2008) Magnetic phase transition in two-phase multiferroics predicted from first principles. Phys Rev B 78: 212406

212 Yamauchi K, Sanyal B, Picozzi S (2007) Interface effects at a half-metal/ferroelectric junction. Appl Phys Lett 91: 062506

213 Zavaliche F, Zheng H, Mohaddes-Ardabili L, et al (2005) Electric field-induced magnetization switching in epitaxial columnar nanostructures. Nano Lett 5: 1793-1796

214 Duan C-G, Velev JP, Sabirianov RF, et al (2008) Tailoring magnetic anisotropy at the ferromagnetic/ferroelectric interface. Appl Phys Lett 92: 122905-122903

215 Bruno P (1989) Tight-binding approach to the orbital magnetic moment and magnetocrystalline anisotropy of transition-metal monolayers. Phys Rev B 39: 865-868

216 van der Laan G (1998) Microscopic origin of magnetocrystalline anisotropy in transition metal thin films. J Phys: Condens Matter 10: 3239-3253

217 Sahoo S, Polisetty S, Duan CG, et al (2007) Ferroelectric control of magnetism in BaTiO$_3$/Fe heterostructures via interface strain coupling. Phys Rev B 76: 092108

218 Rondinelli JM, Stengel M, Spaldin NA (2008) Carrier-mediated magnetoelectricity in complex oxide heterostructures. Nat Nanotech 3: 46-50

219 Weisheit M, Fahler S, Marty A, et al (2007) Electric field-induced modification of magnetism in thin-film ferromagnets. Science 315: 349-351

220 Zhang S (1999) Spin-dependent surface screening in ferromagnets and magnetic tunnel junctions. Phys Rev Lett 83: 640-643

221 Duan C-G, Nan C-W, Jaswal SS, et al (2009) Universality of the surface magnetoelectric effect in half-metals. Phys Rev B 79: 140403

222 Niranjan MK, Duan C-G, Jaswal SS, et al (2010) Electric field effect on magnetization at the Fe/MgO(001) interface. Appl Phys Lett 96: 222504-222503

223 Maruyama T, Shiota Y, Nozaki T, et al (2009) Large voltage-induced magnetic anisotropy change in a few atomic layers of iron. Nat Nanotech 4: 158-161

224 Julliere M (1975) Tunneling between ferromagnetic films. Phys Lett A 54: 225-226

225 Åkerman J (2005) Toward a universal memory. Science 308: 508-510

226 Esaki L, Laibowitz RB, Stiles PJ (1971) Polar switch. IBM Tech Discl Bull 13: 2161

227 Batra IP, Silverman BD (1972) Thermodynamic stability of thin ferroelectric films. Solid State Commun 11: 291-294

228 Bune AV, Fridkin VM, Ducharme S, et al (1998) Two-dimensional ferroelectric films. Nature 391: 874-877

229 Tybell T, Ahn CH, Triscone JM (1999) Ferroelectricity in thin perovskite films. Appl Phys Lett 75: 856-858

230 Rabe KM, Ghosez P (2000) Ferroelectricity in PbTiO$_3$ thin films: A first principles approach. J Electroceram 4: 379-383

231 Sai N, Kolpak AM, Rappe AM (2005) Ferroelectricity in ultrathin perovskite films. Phys Rev B 72: 020101

232 Son J, Cagnon J, Boesch DS, et al (2008) Epitaxial SrTiO$_3$ tunnel barriers on pt/MgO substrates. Appl Phys Exp 1: 061603

233 Rault JE, Ren W, Prosandeev S, et al (2012) Thickness-dependent polarization of strained BiFeO$_3$ films with constant tetragonality. Phys Rev Lett 109: 267601

234 Tenne DA, Bruchhausen A, Lanzillotti-Kimura ND, et al (2006) Probing nanoscale ferroelectricity by ultraviolet raman spectroscopy. Science 313: 1614-1616

235 Kohlstedt H, Pertsev NA, Waser R (2001) Size effects on polarization in epitaxial ferroelectric films and the concept of ferroelectric tunnel junctions including first results. In: Proceedings of the MRS Proceedings, 2001. Cambridge Univ Press: Published

236 Contreras JR, Kohlstedt H, Poppe U, et al (2003) Resistive switching in metal-ferroelectric-metal junctions. Appl Phys Lett 83: 4595-4597

237 Rodriguez Contreras J (2003) Ferroelectric tunnel junctions. thesis. Universität zu Köln,

238 Tsymbal EY, Kohlstedt H (2006) Tunneling across a ferroelectric. Science 313: 181-183

239 Kohlstedt H, Pertsev NA, Rodríguez Contreras J, et al (2005) Theoretical current-voltage characteristics of ferroelectric tunnel junctions. Phys Rev B 72: 125341

240 Luo X, Wang B, Zheng Y (2011) Tunable tunneling electroresistance in ferroelectric tunnel junctions by mechanical loads. ACS Nano 5: 1649-1656

241 Hinsche NF, Fechner M, Bose P, et al (2010) Strong influence of complex band structure on tunneling electroresistance: A combined model and *ab initio* study. Phys Rev B 82: 214110

242 Scott JF (2006) Nanoferroelectrics: Statics and dynamics. J Phys: Condens Matter 18: R361

243 Fang Y-W, Gao Y-C, Gong S-J, et al (2013) Advances in ferroelectric and multiferroic tunnel junctions. Progress in Physics (Chinese) 33: 382-413

244 Valencia S, Crassous A, Bocher L, et al (2011) Interface-induced room-temperature multiferroicity in BaTiO$_3$. Nat Mater 10: 753-758

245 Bocher L, Gloter A, Crassous A, et al (2012) Atomic and electronic structure of the BaTiO$_3$/Fe interface in multiferroic tunnel junctions. Nano Lett 12: 376-382

246 Garcia V, Bibes M, Bocher L, et al (2010) Ferroelectric control of spin polarization. Science 327: 1106-1110

247 Hambe M, Petraru A, Pertsev NA, et al (2010) Crossing an interface: Ferroelectric control of tunnel currents in magnetic complex oxide heterostructures. Adv Funct Mater 20: 2436-2441
248 Yin Y-W, Raju M, Hu W-J, et al (2012) Multiferroic tunnel junctions. Front Phys 7: 380-385
249 Sai N, Meyer B, Vanderbilt D (2000) Compositional inversion symmetry breaking in ferroelectric perovskites. Phys Rev Lett 84: 5636-5639
250 Nakhmanson SM, Rabe KM, Vanderbilt D (2005) Polarization enhancement in two- and three-component ferroelectric superlattices. Appl Phys Lett 87: 102906-102903
251 Wu X, Stengel M, Rabe KM, et al (2008) Predicting polarization and nonlinear dielectric response of arbitrary perovskite superlattice sequences. Phys Rev Lett 101: 087601
252 Lee HN, Christen HM, Chisholm MF, et al (2005) Strong polarization enhancement in asymmetric three-component ferroelectric superlattices. Nature 433: 395-399
253 Yamada H, Kawasaki M, Ogawa Y, et al (2002) Perovskite oxide tricolor superlattices with artificially broken inversion symmetry by interface effects. Appl Phys Lett 81: 4793-4795
254 Gao Y-C, Duan C-G, Tang XD, et al (2013) A first-principles study on the intrinsic asymmetric ferroelectricity of the $SrTiO_3$-$BaTiO_3$-$CaTiO_3$ tricolor superlattice at the nanoscale. J Phys: Condens Matter 25: 165901
255 Wolf SA, Awschalom DD, Buhrman RA, et al (2001) Spintronics: A spin-based electronics vision for the future. Science 294: 1488-1495
256 Bader SD, Parkin SSP (2010) Spintronics. Annu Rev Condens Matter Phys 1: 71-88
257 Martin LW, Chu Y-H, Holcomb MB, et al (2008) Nanoscale control of exchange bias with $BiFeO_3$ thin films. Nano Lett 8: 2050-2055
258 Fechner M, Zahn P, Ostanin S, et al (2012) Switching magnetization by 180° with an electric field. Phys Rev Lett 108: 197206
259 Wang R-Q, Zhu W-J, Ding H-C, et al (2014) Ferroelectric control of in-plane to out-of-plane magnetization switching at poly(vinylidene fluoride)/iron interface. J Appl Phys 115: 043909
260 Winkler R (2003) Spin-orbit coupling effects in two-dimensional electron and hole systems. Berlin ; New York: Springer,
261 Yu AB, Rashba EI (1984) Oscillatory effects and the magnetic susceptibility of carriers in inversion layers. J Phys C: Solid State Phys 17: 6039-6045
262 Nitta J, Akazaki T, Takayanagi H, et al (1997) Gate control of spin-orbit interaction in an inverted $In_{0.53}Ga_{0.47}As$/$In_{0.52}Al_{0.48}As$ heterostructure. Phys Rev Lett 78: 1335-1338
263 Datta S, Das B (1990) Electronic analog of the electro-optic modulator. Appl Phys Lett 56: 665-667
264 Hirsch JE (1999) Spin hall effect. Phys Rev Lett 83: 1834-1837
265 Tse W-K, Das Sarma S (2006) Intrinsic spin hall effect in the presence of extrinsic spin-orbit scattering. Phys Rev B 74: 245309
266 Tse W-K, Fabian J, Žutić I, et al (2005) Spin accumulation in the extrinsic spin hall effect. Phys Rev B 72: 241303
267 Kato YK, Myers RC, Gossard AC, et al (2004) Observation of the spin hall effect in semiconductors. Science 306: 1910-1913
268 Wunderlich J, Kaestner B, Sinova J, et al (2005) Experimental observation of the spin-hall effect in a two-dimensional spin-orbit coupled semiconductor system. Phys Rev Lett 94: 047204
269 Gong S-J, Ding H-C, Zhu W-J, et al (2013) A new pathway towards all-electric spintronics: Electric-field control of spin states through surface/interface effects. Sci China Phys Mech Astron 56: 232-244
270 LaShell S, McDougall BA, Jensen E (1996) Spin splitting of an Au(111) surface state band observed with angle resolved photoelectron spectroscopy. Phys Rev Lett 77: 3419-3422
271 Nuber A, Braun J, Forster F, et al (2011) Surface versus bulk contributions to the Rashba splitting in surface systems. Phys Rev B 83: 165401
272 Bendounan A, Aït-Mansour K, Braun J, et al (2011) Evolution of the Rashba spin-orbit-split shockley state on Ag/Pt(111). Phys Rev B 83: 195427
273 Ast CR, Henk J, Ernst A, et al (2007) Giant spin splitting through surface alloying. Phys Rev Lett 98: 186807
274 Bihlmayer G, Koroteev YM, Echenique PM, et al (2006) The Rashba-effect at metallic surfaces. Surf Sci 600: 3888-3891
275 Park SR, Kim CH, Yu J, et al (2011) Orbital-angular-momentum based origin of Rashba-type surface band splitting. Phys Rev Lett 107: 156803
276 Gong S-J, Duan C-G, Zhu Y, et al (2013) Controlling Rashba spin splitting in Au(111) surface states through electric field. Phys Rev B 87: 035403
277 Krupin O, Bihlmayer G, Döbrich KM, et al (2009) Rashba effect at the surfaces of rare-earth metals and their monoxides. New J Phys 11: 013035
278 Barnes SE, Ieda Ji, Maekawa S (2014) Rashba spin-orbit anisotropy and the electric field control of magnetism. Sci Rep 4: 4105
279 Gong SJ, Li ZY, Yang ZQ, et al (2011) Spintronic properties of graphene films grown on Ni(111) substrate. J Appl Phys 110: 043704
280 Gong SJ, Duan C-G, Zhu Z-Q, et al (2012) Manipulation of magnetic anisotropy of Fe/graphene by charge injection. Appl Phys Lett 100: 122410
281 Ruiz-Díaz P, Dasa TR, Stepanyuk VS (2013) Tuning magnetic anisotropy in metallic multilayers by surface charging: An *ab initio* study. Phys Rev Lett 110: 267203
282 Wang S, Ye H (2014) First-principles calculation of crystalline materials genome: A preliminary study. Chin Sci Bull 59: 1624-1634
283 Curtarolo S, Setyawan W, Wang S, et al (2012) Aflowlib.Org: A distributed materials properties repository from high-throughput ab initio calculations. Comput Mater Sci 58: 227-235
284 Jain A, Hautier G, Moore CJ, et al (2011) A high-throughput infrastructure for density functional theory calculations. Comput Mater Sci 50: 2295-2310
285 Garrity KF, Bennett JW, Rabe KM, et al (2014) Pseudopotentials for high-throughput DFT calculations. Comput Mater Sci 81: 446-452
286 Mueller T, Hautier G, Jain A, et al (2011) Evaluation of tavorite-structured cathode materials for lithium-ion batteries using high-throughput computing. Chem Mater 23: 3854-3862
287 Armiento R, Kozinsky B, Fornari M, et al (2011) Screening for high-performance piezoelectrics using high-throughput density functional theory. Phys Rev B 84: 014103
288 Shenderova OA, Brenner DW, Omeltchenko A, et al (2000) Properties of polycrystalline diamond: Multiscale modeling approach. Mol Simulat 24: 197-207
289 Masanori K (2002) Computational studies of grain boundaries in covalent materials. Modell Simul Mater Sci Eng 10: R31
290 Guyonnet J, Agoritsas E, Bustingorry S, et al (2012) Multiscaling analysis of ferroelectric domain wall roughness. Phys Rev Lett 109: 147601
291 Völker B, Marton P, Elsässer C, et al (2011) Multiscale modeling for ferroelectric materials: A transition from the atomic level to phase-field modeling. Continuum Mech Thermodyn 23: 435-451
292 Zhu W, Xiao D, Liu Y, et al (2014) Picosecond electric field pulse induced coherent magnetic switching in MgO/FePt/Pt(001)-based tunnel junctions: A multiscale study. Sci Rep 4: 4117
293 Yin YW, Burton JD, Kim YM, et al (2013) Enhanced tunnelling electroresistance effect due to a ferroelectrically induced phase transition at a magnetic complex oxide interface. Nat Mater 12: 397-402
294 Dmitrienko VE, Ovchinnikova EN, Collins SP, et al (2014) Measuring the Dzyaloshinskii-Moriya interaction in a weak ferromagnet. Nat Phys 10: 202-206

# Supplementary Information for "First-principles studies of multiferroic and magnetoelectric materials"

Yue-Wen Fang [1], Hang-Chen Ding [1], Wen-Yi Tong[1], Wan-Jiao Zhu[1], Xin Shen[1], Shi-Jing Gong[1*], Xian-Gang Wan [2], Chun-Gang Duan [1,3,*]

[1] Key Laboratory of Polar Materials and Devices, Ministry of Education, East China Normal University, Shanghai 200062, China;

[2] Department of Physics and National Laboratory of Solid State Microstructures, Nanjing University, Nanjing 210093, China;

[3] National Laboratory for Infrared Physics, Chinese Academy of Sciences, Shanghai 200083, China

[*]Corresponding authors (email: sjgong@ee.ecnu.edu.cn, wxbdcg@gmail.com)

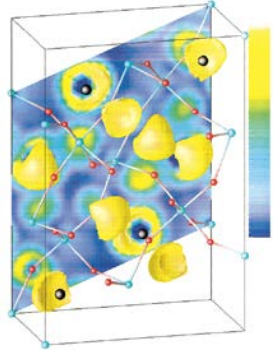

**Fig. S1** (Color online) Valence electron localization functions for monoclinic BMO. The blue end of the scale bar corresponds to no electron localization while the white end corresponds to a complete localization. Reproduced with permission from [1]. Copyright 2001, American Chemical Society

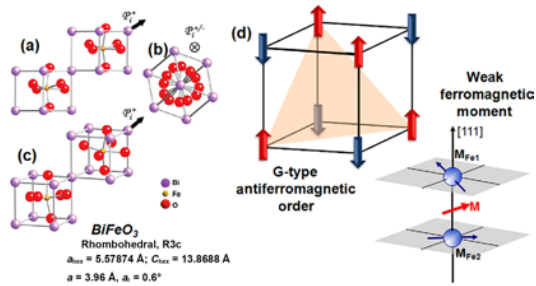

**Fig. S2** (Color online) Structure of BFO shown **a** the pseudocubic-[110] direction, **b** the pseudocubic-[111] polarization direction, and **c** a general three dimensional view of the structure. **d** G-type antiferromagnetic ordering and spiral magnetic ordering. Reproduced with permission from [2]. Copyright 2008, Institute of Physics

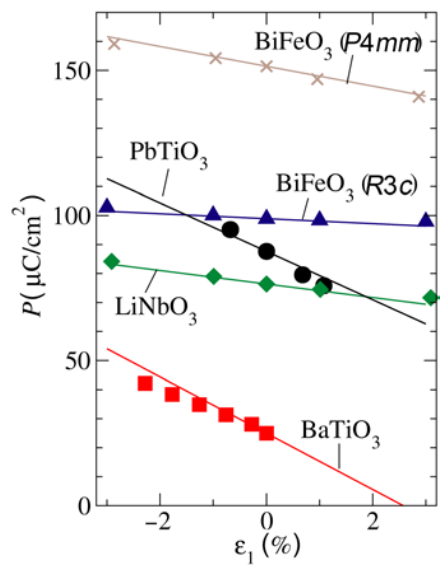

**Fig. S3** (Color online) Spontaneous polarization $P$ as a function of epitaxial strain $\varepsilon_1$. Symbols correspond to results from first-principles calculations for strained unit cells, lines are obtained from the calculated bulk linear response functions. Reproduced with permission from [3]. Copyright 2005, American Physical Society

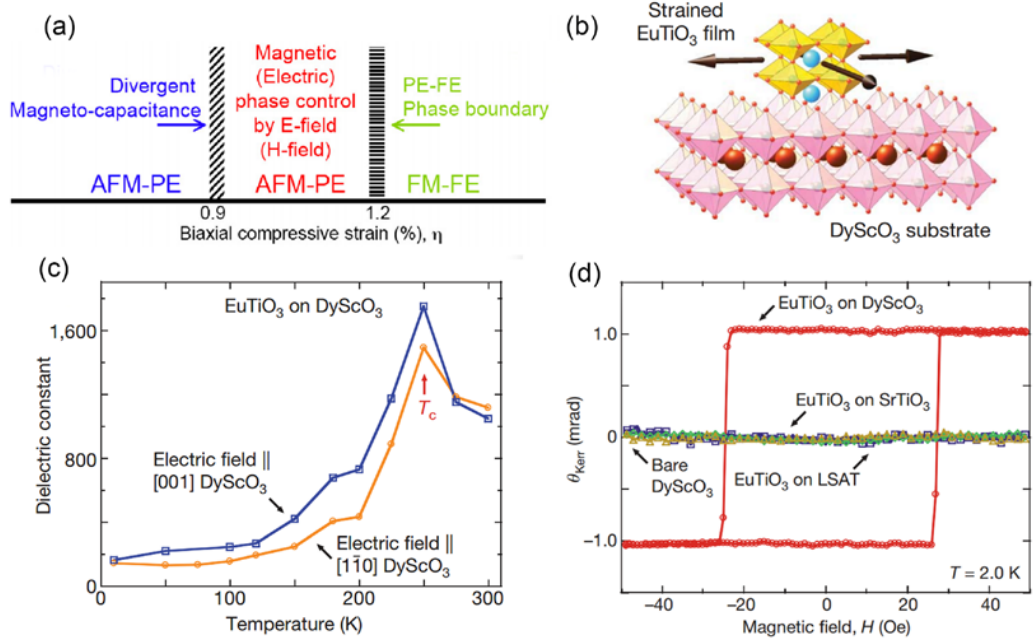

**Fig. S4** (Color online) **a** First-principles compressive epitaxial strain phase diagram of EuTiO$_3$. **b** Schematic of epitaxially strained thin-film EuTiO$_3$ on the DyScO$_3$ substrate. **c** Dielectric constant versus temperature by far-infrared reflectance spectroscopy showing that 100-nm-thick EuTiO$_3$-on-DyScO$_3$ film is ferroelectric below $T_c \approx 250$ K. **d** Magneto-optic Kerr effect measurements at 2.0 K of EuTiO$_3$ on different substrates showing that EuTiO$_3$ on DyScO$_3$ is ferromagnetic below the temperature. Reproduced with permission from [4, 5]. Copyright 2006, American Physical Society and Copyright 2010 Nature Publishing Group, respectively

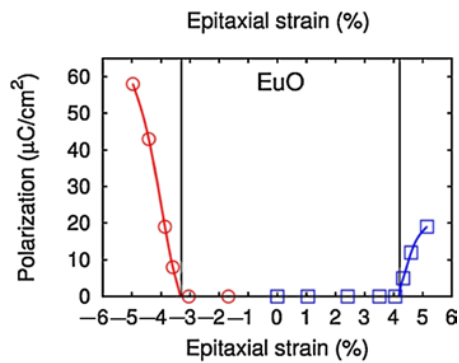

**Fig. S5** (Color online) Amplitude of the spontaneous polarization of EuO with respect to epitaxial strain. The paraelectric-to-ferroelectric transition will occur when the biaxial compressive strain is larger than 3.3 % or the biaxial tensile strain is larger than 4.2 %. Reproduced with permission from [6]. Copyright 2010, American Physical Society

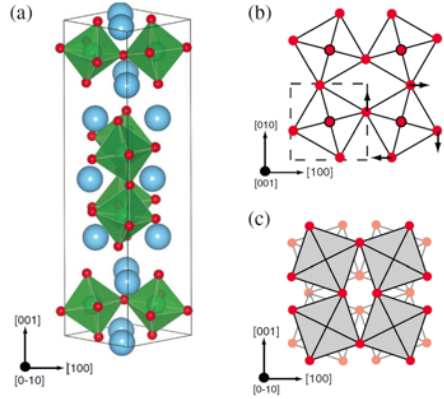

**Fig. S6** (Color online) Ca$_3$Mn$_2$O$_7$ structure and rotation distortions. **a** The $A2_1am$ ferroelectric ground state structure. Large spheres correspond to Ca ions. **b**, **c** Schematic of two different octahedron rotation modes $X_2^+$ and $X_3^-$, all axes refer to the coordinate system of the $I4/mmm$ parent structure. Reproduced with permission from [7]. Copyright 2011, American Physical Society

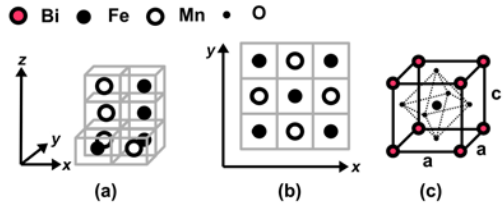

**Fig. S7** (Color online) **a** BFO-BMO atomic-scale checkerboard. **b** Top view of the atomic-scale checkerboard. **c** Perovskite cell. Dashed lines show an oxygen octahedron surrounding the B-site (Fe or Mn) cation. Reproduced with permission from [8]. Copyright 2010, American Physical Society

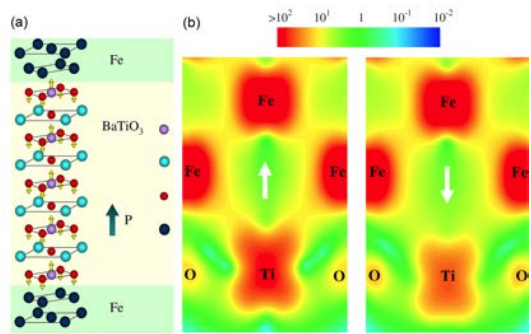

**Fig. S8** (Color online) **a** Atomic structure of Fe/BTO multilayer. **b** Minority-spin charge density at the Fe/BTO interface for two opposite polarizations in BTO. Reproduced with permission from [9]. Copyright 2006, American Physical Society

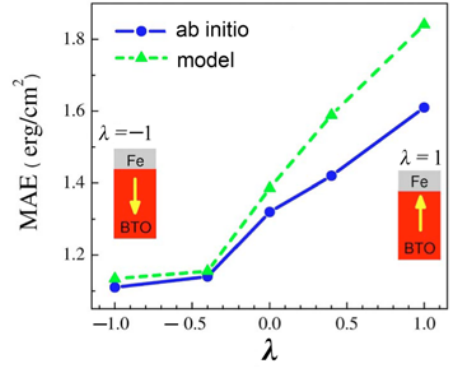

**Fig. S9** (Color online) MAE as a function of the polarization factor $\lambda$, here $\lambda = 1$ and $-1$ corresponds the polarization up and down. Reproduced with permission from [10]. Copyright 2008, American Institute of Physics

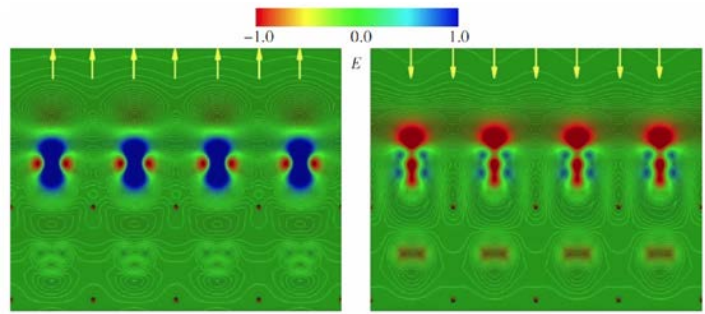

**Fig. S10** (Color online) Surface magnetoelectric effect of Fe film. Left: increase of the magnetization when the electric field is pointed away from the surface. Right: decrease of the magnetization when the electric field is pointed towards the surface. Reproduced with permission from [11]. Copyright 2008, American Physical Society

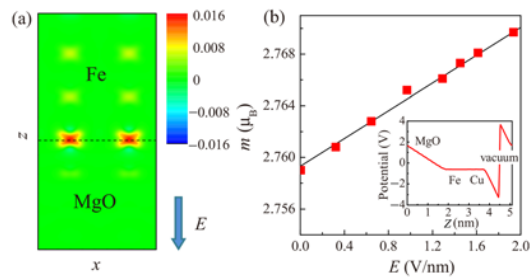

**Fig.S11** (Color online) **a** Induced spin density $\Delta\sigma=\sigma(E)-\sigma(0)$, in units of e/Å$^3$, projected to the *x-z* or (010) plane around the Fe/MgO interface under the influence of electric field $E = 1.0$ V/nm in MgO. The dashed line indicates the interfacial Fe monolayer at the Fe/MgO interface. **b** Magnetic moment (in units of $\mu_B$) of Fe at Fe/MgO interface as a function of the electric field in the MgO. The inset shows the calculated electrostatic potential across the supercell due to the applied electric field $E = 4$ V/nm. Reproduced with permission from [12]. Copyright 2010, American Institute of Physics

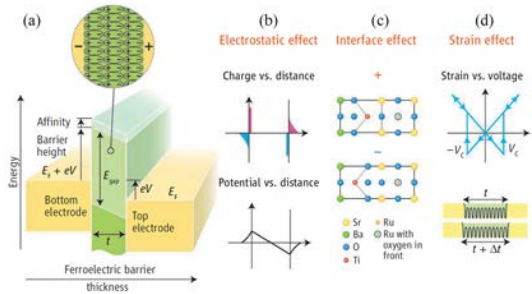

**Fig. S12** (Color online) A ferroelectric tunnel junction. Schematic diagram of a tunnel junction, which consists of two electrodes separated by a nanometer-thick ferroelectric barrier layer. ($E_{gap}$ is the energy gap. $E_F$ is the Fermi energy, $V$ is the applied voltage, $V_c$ is the coercive voltage, $t$ is the barrier thickness, and $\Delta t$ is the thickness variation under an applied field. Reproduced with permission from [13]. Copyright 2006, Science

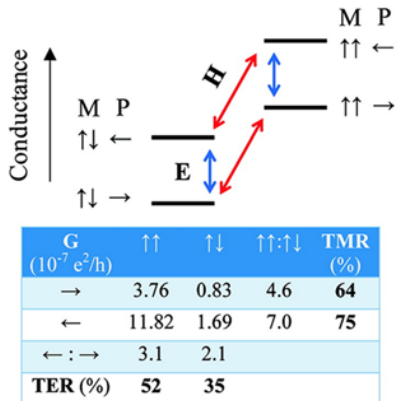

**Fig. S13** (Color online) Conductance of the SRO/BTO/SRO MFTJ. The four conductance states are distinguished by polarization in the barrier pointing to the left (←) or right (→) and magnetization of the electrodes being parallel (↑↑) or antiparallel (↓↓). Conductance values are given per transverse area of the unit cell. The diagram on the top shows schematically the four resistance states that can be controlled by electric ($E$) and magnetic ($H$) fields. Reproduced with permission from [14]. Copyright 2009, American Chemical Society

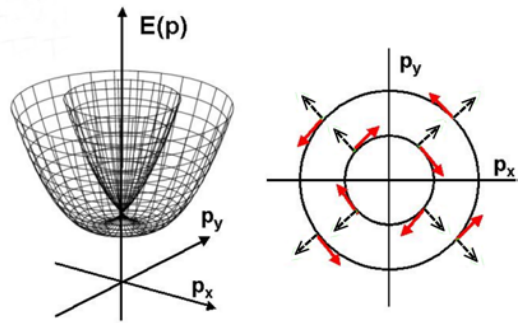

**Fig. S14** (Color online) Two-dimensional electronic eigenstates in a Rashba spin-orbit coupled system are labeled by momentum (dashed arrows). For each momentum the two eigenspinors point in the azimuthal direction (solid arrows). Reproduced with permission from [15]. Copyright 2004,

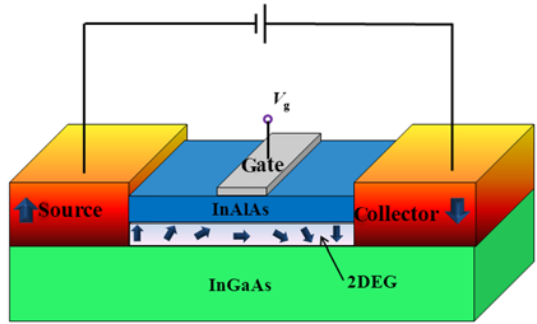

**Fig. S12** (Color online) Schematic diagram of the Datta-Das type spin field effect transistor. Reproduced with permission from [16]. Copyright 2013, Springer

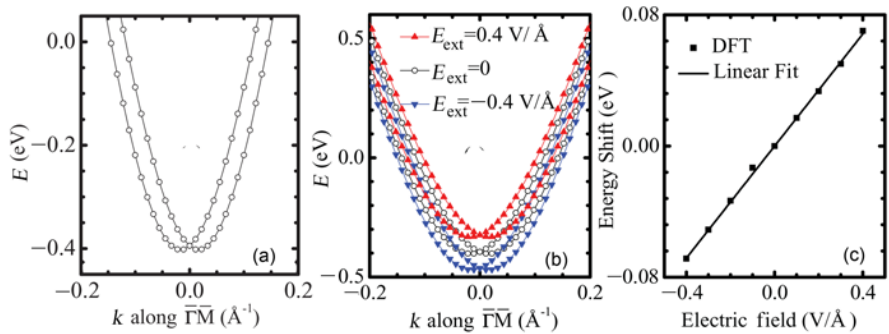

**Fig. S13** (Color online)  **a** Rashba splitting bands of 22-layer Au(111) along $M$, **b** the surface Rashba splitting bands under the electric fields $E_{ext}$ = 0 (open circles), 0.4 V/Å (up-triangles), and −0.4 V/Å (down-triangles), **c** Electric-field dependence of the band energy shift (squares). The solid line is a linear fit to the calculated data. Reproduced with permission from [17]. Copyright 2013, American Physical Society

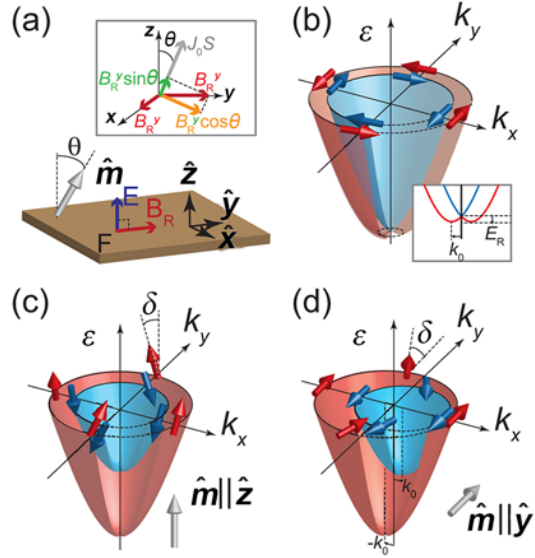

**Fig. S17** (Color online) **a** The electric field $E$ is perpendicular to the ferromagnet surface while the order parameter direction $m$, is defined by the angle $\theta$ relative to $z$. Whatever the direction of $k$, the Rashba magnetic field $B_R$ of direction $k{\times}E$ lies in the $x$-$y$ plane. **b** The Rashba split bands of a nonmagnetic metal. The two Fermi sheets emerge from a ''Dirac point'' near the bottom of the illustration. For the magnetic case the two Fermi sheets are disconnected. **c** For a perpendicular $m$ the electron spins make a constant angle $\delta$ to the vertical such that the projection is as in **b**. **d** Same but for $m$ parallel to the plane. With $m$ along the $y$-direction the majority and minority Fermi seas shift along the $x$-axis in opposite directions. The tilt of the spin relative to $m$ is no longer a constant being zero along the $x$-axis and a maximum along the $y$-axis. Reproduced with permission from [18]. Copyright 2014, Nature Publishing Group

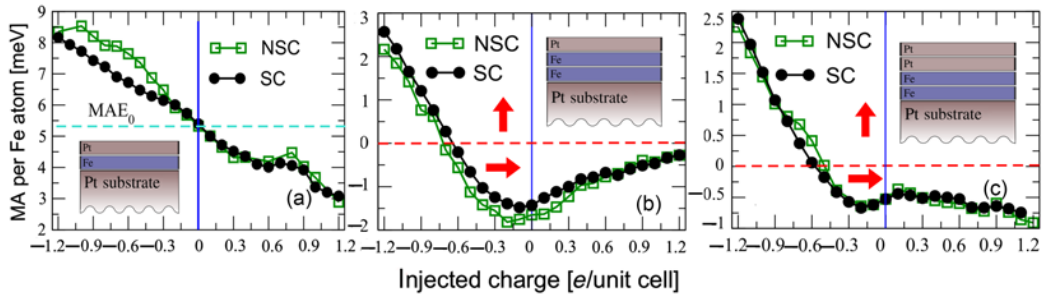

**Fig. S18** (Color online) Calculated self-consistent (SC) and nonself-consistent (NSC) magnetic anisotropy energy (in meV/per magnetic atom) for **a** Pt/Fe/Pt(001), **b** Pt/Fe$_2$/Pt(100) and **c** Pt$_2$/Fe$_2$/Pt(100) as a function of the injected charge (holes). The charge-doping scale (in units of e/unit cell) is referred to the neutral system. Positive (negative) values stand for an excess (lack) of valence electrons. The dashed lines are only to guide the eye to find the MA in the neutral system. Positive (negative) MA values stand for an out (in-plane) axis of magnetization. Reproduced with permission from [19]. Copyright 2013, American Physical Society

# References

1. Seshadri R, Hill NA (2001) Visualizing the role of bi 6s "lone pairs" in the off-center distortion in ferromagnetic $BiMnO_3$. Chem Mater 13: 2892-2899
2. Martin LW, Crane SP, Chu YH, et al (2008) Multiferroics and magnetoelectrics: Thin films and nanostructures. J Phys: Condens Matter 20: 434220
3. Ederer C, Spaldin NA (2005) Effect of epitaxial strain on the spontaneous polarization of thin film ferroelectrics. Phys Rev Lett 95: 257601
4. Fennie CJ, Rabe KM (2006) Magnetic and electric phase control in epitaxial $EuTiO_3$ from first principles. Phys Rev Lett 97: 267602
5. Lee JH, Fang L, Vlahos E, et al (2010) A strong ferroelectric ferromagnet created by means of spin-lattice coupling. Nature 466: 954-U972
6. Bousquet E, Spaldin NA, Ghosez P (2010) Strain-induced ferroelectricity in simple rocksalt binary oxides. Phys Rev Lett 104: 037601
7. Benedek NA, Fennie CJ (2011) Hybrid improper ferroelectricity: A mechanism for controllable polarization-magnetization coupling. Phys Rev Lett 106: 107204
8. Palova L, Chandra P, Rabe KM (2010) Magnetostructural effect in the multiferroic $BiFeO_3$-$BiMnO_3$ checkerboard from first principles. Phys Rev Lett 104: 037202
9. Duan C-G, Jaswal SS, Tsymbal EY (2006) Predicted magnetoelectric effect in Fe/$BaTiO_3$ multilayers: Ferroelectric control of magnetism. Phys Rev Lett 97: 047201
10. Duan C-G, Velev JP, Sabirianov RF, et al (2008) Tailoring magnetic anisotropy at the ferromagnetic/ferroelectric interface. Appl Phys Lett 92: 122905-122903
11. Duan C-G, Velev JP, Sabirianov RF, et al (2008) Surface magnetoelectric effect in ferromagnetic metal films. Phys Rev Lett 101: 137201
12. Niranjan MK, Duan C-G, Jaswal SS, et al (2010) Electric field effect on magnetization at the Fe/MgO(001) interface. Appl Phys Lett 96: 222504-222503
13. Tsymbal EY, Kohlstedt H (2006) Tunneling across a ferroelectric. Science 313: 181-183
14. Velev JP, Duan C-G, Burton JD, et al (2009) Magnetic tunnel junctions with ferroelectric barriers: Prediction of four resistance states from first principles. Nano Lett 9: 427-432
15. Sinova J, Culcer D, Niu Q, et al (2004) Universal intrinsic spin hall effect. Phys Rev Lett 92: 126603
16. Gong S-J, Ding H-C, Zhu W-J, et al (2013) A new pathway towards all-electric spintronics: Electric-field control of spin states through surface/interface effects. Sci China Phys Mech Astron 56: 232-244
17. Gong S-J, Duan C-G, Zhu Y, et al (2013) Controlling Rashba spin splitting in Au(111) surface states through electric field. Phys Rev B 87: 035403
18. Barnes SE, Ieda Ji, Maekawa S (2014) Rashba spin-orbit anisotropy and the electric field control of magnetism. Sci Rep 4: 4105
19. Ruiz-Díaz P, Dasa TR, Stepanyuk VS (2013) Tuning magnetic anisotropy in metallic multilayers by surface charging: An *ab initio* study. Phys Rev Lett 110: 267203